\RequirePackage{fix-cm}
\documentclass[smallextended]{svjour3}       
\smartqed  
\usepackage{natbib}
\usepackage{graphicx}
\usepackage[most]{tcolorbox}
\usepackage{listings}
\usepackage{xcolor}
\usepackage{subcaption}
\usepackage{tabularx}
\usepackage{longtable}
\usepackage{booktabs}
\usepackage{wrapfig}
\usepackage{framed}
\newcommand{\highlight}[1]{\begin{framed}%
  \noindent#1
\end{framed}}

\definecolor{backcolor}{rgb}{0.95,0.95,0.92}
\colorlet{punct}{red!60!black}
\definecolor{delim}{RGB}{20,105,176}
\colorlet{numb}{magenta!60!black}

\lstdefinestyle{mystyle}{
  language=Python,
  basicstyle=\ttfamily\footnotesize,
  keywordstyle=\color{delim},
  stringstyle=\color{punct},
  commentstyle=\color{numb},
  breakatwhitespace=true,
  breaklines=true,
  keepspaces=true,
  showspaces=false,
  showstringspaces=false,
  showtabs=false,
  tabsize=2,
  stepnumber=1,
  numbersep=3pt,
  captionpos=b,
  backgroundcolor=\color{backcolor},
}
\begin{document}

\title{Understanding Feedback Mechanisms in Machine Learning Jupyter Notebooks}


\author{Arumoy Shome\and
Lu{\'\i}s Cruz\and
  Diomidis Spinellis\and
  Arie van Deursen
}


\institute{A. Shome \at
  Delft University of Technology\\
  \email{a.shome@tudelft.nl}
  \and
  L. Cruz \at
  Delft University of Technology\\
  \email{l.cruz@tudelft.nl}
  \and
  D. Spinellis \at
  Delft University of Technology\\
  \email{d.spinellis@tudelft.nl}
  \and
  A. V. Deursen \at
  Delft University of Technology\\
  \email{arie.vandeursen@tudelft.nl}
}

\date{Received: date / Accepted: date}

\maketitle

\begin{abstract}
  The machine learning development lifecycle is characterized by iterative and exploratory processes that rely on feedback mechanisms to ensure data and model integrity. Despite the critical role of feedback in machine learning engineering, no prior research has been conducted to identify and understand these mechanisms. To address this knowledge gap, we mine 297.8 thousand Jupyter notebooks and analyse 2.3 million code cells. We identify three key feedback mechanisms---assertions, print statements and last cell statements---and further categorize them into implicit and explicit forms of feedback. Our findings reveal extensive use of implicit feedback for critical design decisions and the relatively limited adoption of explicit feedback mechanisms. By conducting detailed case studies with selected feedback instances, we uncover the potential for automated validation of critical assumptions in ML workflows using assertions. Finally, this study underscores the need for improved documentation, and provides practical recommendations on how existing feedback mechanisms in the ML development workflow can be effectively used to mitigate technical debt and enhance reproducibility.
\end{abstract}


\section{Introduction}

The machine learning development lifecycle (MLDL) encompasses a multifaceted series of entangled stages. Significant effort is invested in gathering and assembling the required data from diverse sources. The data then undergoes cleaning and preprocessing to ensure high quality and usability, in the subsequent stages. MLDL is dominated by the model development phase which is highly experimental and iterative. This stage involves exploratory data analysis (EDA) to uncover underlying patterns, fitting the data to various models, and rigorously analysing the results. Often, this stage may lead to additional rounds of feature engineering, where new attributes are created based on insights gained from initial models to enhance model performance. Practitioners may also engage in comparing and contrasting different models to identify the most effective approach, followed by fine-tuning selected models to optimize performance. The final model is then deployed into production where it is continuously monitored, and periodically retrained to adapt to new data or changing conditions~\citep{haakman2021ai,amershi2019software,sculley2015hidden}.

The MLDL is fundamentally iterative, closely resembling Agile software engineering principles, which emphasize incorporating early feedback and working in small, manageable iterations~\citep{betz2018managing}. Similarly, CRISP-DM---the dominant process model in data mining---outlines a cyclical process that evolves from understanding the business objectives and data, to data preparation, modelling, evaluation and finally deployment and monitoring~\citep{martinez-plumed2021crisp-dm}. Both systems advocate for a dynamic, iterative approach that enhances adaptability and effectiveness in dealing with complex, changing systems like those encountered in ML projects.

Computational notebooks---specifically Jupyter notebooks\footnote{https://jupyter.org/}---have been ubiquitously adopted by the ML community for developing ML enabled systems~\citep{pimentel2019large-scale,quaranta2021kgtorrent,psallidas2019data,perkel2018why}. Jupyter notebooks provide an interactive computing environment that allow users to break the development of complex workflows into smaller, more manageable code chunks. Each notebook is composed of a series of code cells, which can be executed independently in a read-eval-print loop (REPL) style of interactive software development. This REPL approach allows users to write a piece of code, run it, and immediately see the results thus facilitating an incremental and iterative style of software development. This modular structure of notebooks provides users the flexibility to experiment with different approaches and algorithms. Users can test hypotheses, debug issues, and make incremental changes with immediate feedback. The ability to visualize data and results inline and document the process in a narrative form further enhances their utility~\citep{kery2018story,head2019managing,rule2018exploration,chattopadhyay2020whats}.

Notebook users can obtain feedback using implicit or explicit mechanisms. Implicit feedback mechanisms require manual validation by the user. For instance, a common prerequisite for all ML models is that the features in the dataset are all numerical. This can be verified with a print statement in a code cell and manually checking the output. This code statement however, will continue to execute even if the data type of a feature changes in subsequent batches of training data.

In contrast, explicit feedback mechanisms document the expected conditions of the code and halt execution if these conditions are not met. For instance, the implicit validation method mentioned above can be replaced with an explicit assertion: \lstinline{assert all(df[col].dtype in ['int64', 'float64'] for col in df.columns), "Not all columns are of numerical type"}. This assert statement will stop the execution if the condition is not satisfied and provide immediate feedback with a clear message explaining the issue.

As machine learning pipelines developed within Jupyter notebooks become business-critical, they are frequently transitioned into automated scripts for production environments~\citep{kery2018story,rule2018exploration}. In these contexts, it is crucial to shift from implicit feedback mechanisms to explicit ones. This transition reduces the reliance on manual verification, thereby enhancing the robustness and maintainability of ML pipelines.

To the best of our knowledge, no prior studies have been conducted to understand how feedback mechanisms are used by practitioners when developing ML pipelines inside Jupyter notebooks. To address this gap, we conduct a series of empirical studies to understand the role of various feedback mechanisms within Jupyter notebooks. We focus on implicit form of feedback obtained from print statements and output generated by the last statement of code cells, as well as explicit form of feedback obtained from assertions. The research questions along with the contributions of this study are presented below.

\begin{description}
  \item[RQ1.] \textbf{What feedback mechanisms are employed in the development of ML Jupyter notebooks?}

    We mine 297.8 thousand Jupyter notebooks written in Python from GitHub and Kaggle. We analyse 2.3 million code cells obtained from the notebooks and identify three key feedback mechanisms used in ML notebooks---explicit feedback from assertions and implicit feedback from print and last cell statements.

    We create a public dataset of 89.6 thousand assertions, 1.4 million print statements and 1 million last cell statements. We perform a descriptive and lexical analysis of the dataset and report the key characteristics of feedback mechanisms used in ML Jupyter notebooks. The dataset along with the replication package is shared publicly under the Creative Commons Attribution (CC-BY) licence\footnote{https://doi.org/10.6084/m9.figshare.26372140.v1}.

  \item[RQ2.] \textbf{How is explicit feedback from assert statements used to validate ML code written in Jupyter notebooks?}

    To gain a deeper understanding of Python assertions written in Jupter notebooks, we conduct individual case-studies with 82 assert statements. We analyse the assertion along with the surrounding code to understand its purpose. Additionally, we analyse the entire code cell, the previous and next cells along with the purpose of the notebook to bring in rich context.

  \item[RQ3.] \textbf{How is implicit feedback from print statements and last cell statements used when writing ML code in Jupyter notebooks?}

    Similarly, we conduct individual case-studies with 44 print statements and 27 last cell statements. During each case-study, we analyse the code statement along with the raw output to understand its purpose.
\end{description}

Our findings indicate that implicit feedback mechanisms are extensively utilized for making critical design and implementation decisions during the MLDL. While manual validation of these outputs is common, it is also prone to inconsistencies thus highlighting the need for automated validation mechanisms. Assertions are identified as a potential solution, capable of verifying data distributions and ensuring model performance metrics meet predefined thresholds. However, the study also notes the limited use of assertions, with only 8.5\% of the notebooks analysed incorporating them, suggesting a significant gap in current testing practices within the ML community. The findings underscore the necessity for developing ML-specific testing tools that are deeply integrated into the ML development workflow to ensure consistent and reliable validation processes.

\section{Background}\label{sec:background}

This section introduces relevant prior concepts for this study.

\subsection{REPLs, Interactive Computing and Project Jupyter}

A Read-Eval-Print Loop (REPL) is an interactive programming environment that takes single user inputs (Read), executes them (Eval), returns the result to the user (Print) and then waits for the next command (Loop). REPLs offer an immediate feedback loop for quickly testing and debugging code snippets. The Python standard distribution includes a Python REPL with a command line interface to execute Python code in real-time. IPython---a more advanced alternative to the Python REPL---extends the capabilities of the standard Python shell with additional features such as powerful introspection, additional shell syntax, tab completion, and command history. Jupyter notebooks (and its successor Jupyter lab) leverage the IPython kernel to execute Python code and extends its capabilities with a graphical user interface (GUI). Resembling the literate programming paradigm~\citep{knuth1984literate}, Jupyter notebooks offer a web-based interface where users can create and share documents that contain live code, equations, visualizations, and narrative text.

\subsection{Anatomy of Jupyter Notebooks}

\begin{figure}
  \centering
  \includegraphics[width=\linewidth]{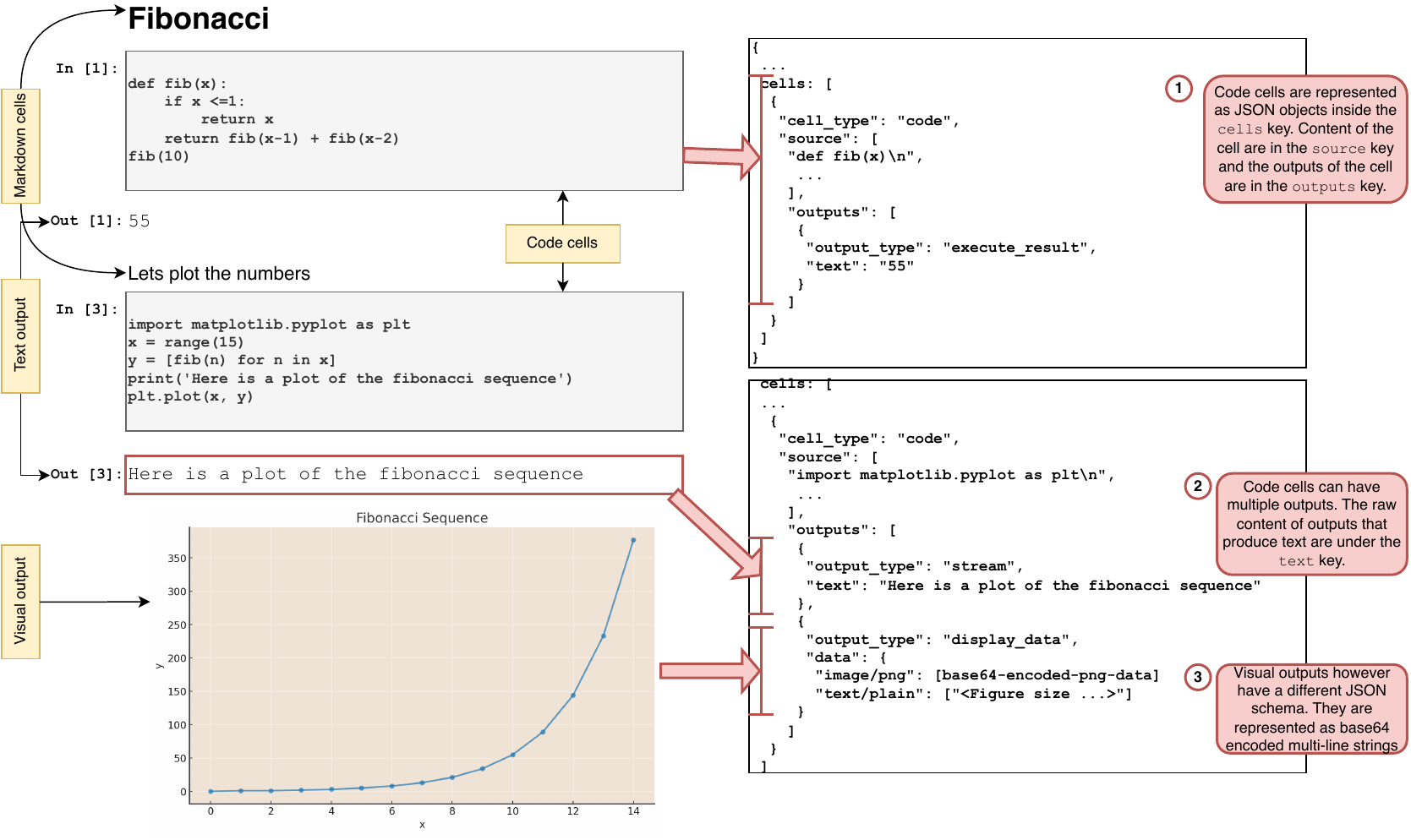}
  \caption{Example Jupyter notebook with code and markdown cells, adapted from~\citep[Figure~1]{pimentel2019large-scale}.}
  \label{fig:nb}
\end{figure}

Figure~\ref{fig:nb} shows an example Jupyter notebook with markdown and code cells. The data comprised within Jupyter notebooks is stored using the JSON schema specified by \emph{nbformat}\footnote{https://nbformat.readthedocs.io/en/latest/index.html}, and thus can be parsed programmatically. A notebook consists of \emph{cells} which can be of three types---``markdown'', ``code'' and ``raw''. All cells of a notebook are stored as a list under the top-level \lstinline[language={}]$cells$ key in the JSON representation. Within this list, each cell is represented as a JSON object. The contents of the cell are stored as a list of strings inside the \lstinline[language={}]$source$ key.

Only code cells in Jupyter notebooks can generate outputs. As shown in Figure~\ref{fig:nb}, outputs can come from three main sources in the source code---print statements, the last statement in the code cell, and statements that create a visualization. The outputs produced by code cells are stored as JSON objects under the \lstinline[language={}]$outputs$ key in the corresponding JSON representation of the notebook. Text outputs (such as those produced by the last statement of a cell or print statements) are stored under the \lstinline[language={}]$text$ key. Visual outputs have a sightly different JSON schema and are stored under the \lstinline[language={}]$data$ key. The image itself is represented as a base64 encoded string which is stored inside the \lstinline[language={}]$data$ object, under the \lstinline[language={}]$image/png$ key.

\subsection{Python Abstract Syntax Tree}

The Python \lstinline{ast} module enables the parsing, manipulation, and analysis of Python source code by converting it into an abstract syntax tree (AST)\footnote{https://docs.python.org/3/library/ast.html}. The \lstinline{ast} module can parse source code strings into AST objects, traverse and modify the tree and compile it back into executable code, providing a powerful tool for dynamic code manipulation. In Python, \lstinline{assert} statements are represented in the AST by the \lstinline{Assert} class. The \lstinline{Assert} node includes the \lstinline{test} attribute---which contains the condition being asserted---and the \lstinline{msg} attribute---which holds an optional message displayed if the assertion fails. Print statements are treated as a call to the built-in \lstinline{print} function. Function calls are represented in the AST by the \lstinline{Call} class. The \lstinline{Call} node includes the \lstinline{func} attribute representing the \lstinline{print} function, and \lstinline{args} which lists the arguments passed.

\section{Methodology}

\subsection{Data Collection}\label{sec:data-collect}

\begin{figure}
  \centering
  \includegraphics[width=0.75\linewidth]{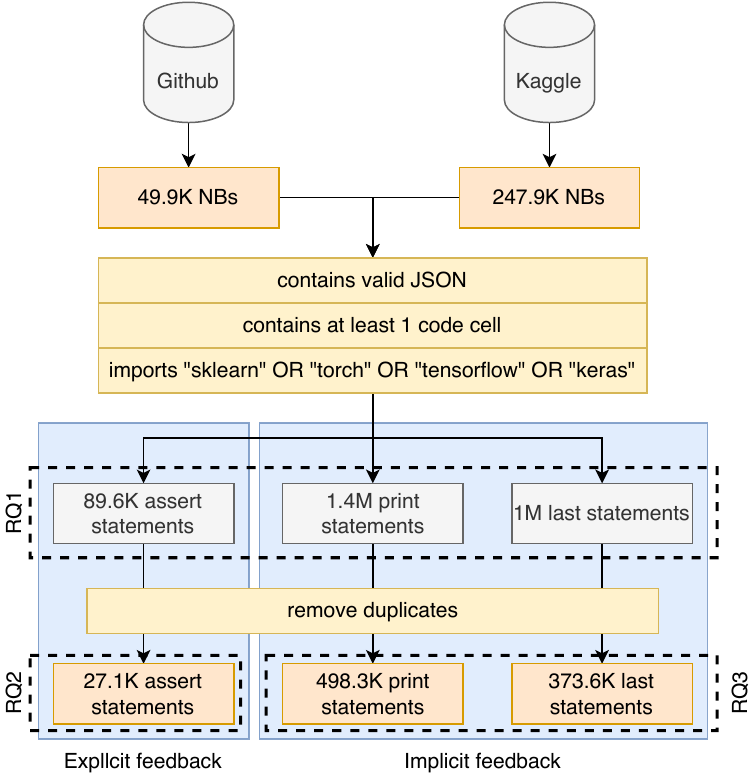}
  \caption{Overview of data collection methodology used in this study}
  \label{fig:data-collection}
\end{figure}

Figure~\ref{fig:data-collection} illustrates the data collection process employed in this study to gather Jupyter notebooks written in Python from GitHub and Kaggle. We used GitHub's advanced search syntax\footnote{https://docs.github.com/en/search-github/searching-on-github/searching-code}, with the following search query: \lstinline[language={}]$language:"Jupyter Notebook"$ to mine 49.9 thousand Jupyter notebooks from public repositories on GitHub\footnote{Collected on June 22, 2023}. We used the pre-existing dataset KGTorrent~\citep{quaranta2021kgtorrent} for Kaggle since it does not support advanced code-based search like GitHub. To the best of our knowledge, KGTorrent is the largest dataset of Python Jupyter notebooks obtained from Kaggle, consisting of 247.9 thousand notebooks. Consequently, we started with 283 GB of data comprising 297.8 thousand Python Jupyter notebooks.

Since the focus of this study is on the analysis of Python code, we only include notebooks that have a valid underlying JSON structure and contain at least one code cell. Finally, to focus on Python code written specifically for ML projects, we only include notebooks that import at least one of the following popular ML libraries---Scikit Learn~\citep{pedregosa2011scikit-learn}, Pytorch~\citep{paszke2017automatic}, Tensorflow~\citep{abadi2015tensorflow} and Keras~\citep{chollet2015keras}. The libraries are selected based on the import analysis conducted by \citet{psallidas2019data} on 6 million Jupyter notebooks collected from GitHub. We use the Python \lstinline{ast} module to extract the assert, print and last statements from all code cells. Prior to conducting the analysis for RQ2 and RQ3, we remove duplicate data points resulting in a data set of 27.1 thousand assertions, 498.3 thousand print statements and 373.6 thousand last cell statements.

\subsection{Case Studies}

For RQ2 and RQ3, we allocated a fixed time resource of 200 hours to conduct the case study analysis of all candidate assertions, print statements and last cell statements. To surface interesting candidates for the case studies, we apply text processing techniques as described below.

The assertions, print statements and last cell statements are first tokenized---special characters and alphanumeric words shorter than two characters are removed. Two stop words namely ``assert'' and ``print'' are removed since they appear in all assertions and print statements respectively. The term frequency (TF) for all tokens is calculated and then normalized using their inter-document frequency (IDF) such that tokens that appear less frequently are assigned a higher value. We apply stratified random sampling to identify the candidates for the case study analysis. The subgroups are created by adding TF-IDF of the tokens in each candidate to produce an aggregate value. The candidates are then divided into quartiles based on the aggregate value. A candidate is randomly drawn from each bin and analysed as an in-depth case study. The analysis is stopped when the time resource is exhausted.

During each case study, we analyse the code of the candidate to understand its purpose. Additionally, we analyse the entire code cell, the previous cell, next cell and the notebook's purpose to bring in rich context. For case studies where the first author was unable to determine the purpose of the candidate, it was additionally analysed by the second and third authors. Further discussions were held with all authors until a consensus was reached.

\section{Results}

\begin{figure}
  \centering
  \includegraphics[width=0.75\textwidth]{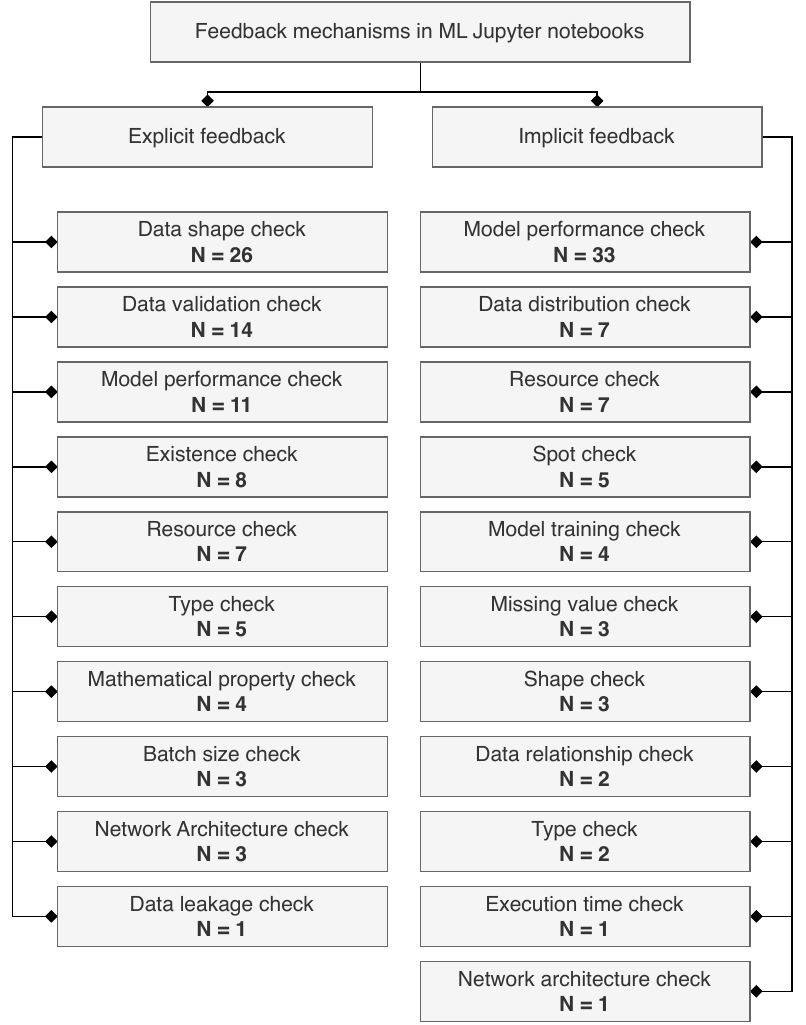}
  \caption{Overview of various feedback mechanism groups identified in this study.}
  \label{fig:taxonomy}
\end{figure}

The results of the study are presented in this section. For the case studies conducted for RQ2 and RQ3, we group the candidates based on their similarity to each other. This is done only to structure the paper in a meaningful manner and for ease of reporting the results. Figure~\ref{fig:taxonomy} provides an overview of the various feedback mechanism groups identified in this study with additional details provided in Section~\ref{sec:result-explicit} and Section~\ref{sec:result-implicit}. Throughout the text, we reference the candidate \textbf{(A)}ssertions, \textbf{(P)}rint statements and \textbf{(L)}ast cell statements using their unique identifiers as examples. These can be viewed on our online appendix\footnote{https://arumoy.me/shome2023notebook}.

\subsection{(RQ1) What feedback mechanisms are employed in the development of ML Jupyter notebooks?}~\label{sec:result-analysis}

25 thousand (8.5\%) notebooks contained at least one assertion. From the 8.5\% notebooks, we collected 89.6 thousand assertions. 4.8 thousand assertions (5\%) used methods from external testing libraries. 85 thousand assertions (95\%) however were written using the \lstinline{assert} statement provided by the built-in Python standard library. We removed duplicate assertions for the remainder of the analysis, thus resulting in 27.1 thousand assertions.

\begin{figure}
  \centering
  \includegraphics[width=0.5\textwidth]{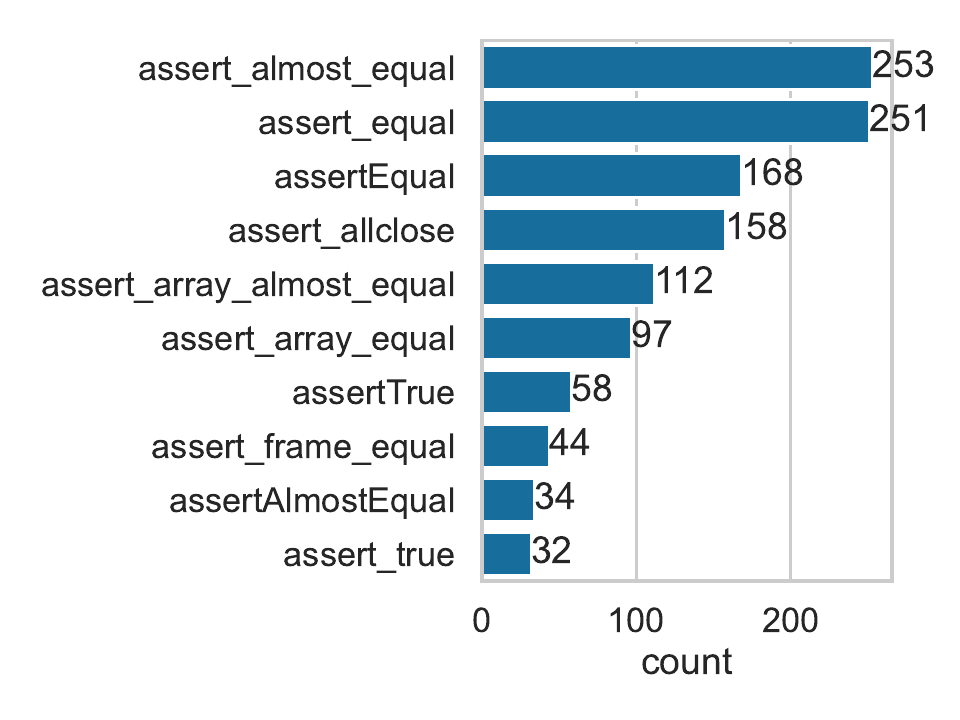}
  \caption{10 most common methods used in assertions written using external testing libraries.}
  \label{fig:other-test-methods}
\end{figure}

Figure~\ref{fig:other-test-methods} presents the 10 most common methods used in the 5\% assertions written using external testing libraries. We manually tracked the source of the methods to the numpy, pandas, torch and unittest libraries. The most frequently used methods include \lstinline{assert_almost_equal} and \lstinline{assert_equal}, each with over 250 instances. Methods such as \lstinline{assert_almost_equal}, \lstinline{assert_array_almost_equal} and \lstinline{AssertAlmostEqual} accommodate the stochastic nature of ML, where results can vary slightly due to inherent randomness in data processing and model training. We also find methods such as \lstinline{assert_array_equal} and \lstinline{assert_frame_equal} that are specifically tailored for comparing high-dimensional data structures such as numpy arrays and pandas dataframes.

\highlight{\textbf{Finding 1.1} 25 thousand (8.5\%) notebooks contained at least one assertion. 95\% of the 89.6 thousand assertions were written using the built-in Python \lstinline{assert} statement and 5\% using external testing libraries. Methods such as \lstinline{assert_almost_equal} and \lstinline{assert_array_equal} from the numpy and pandas libraries were frequently used to accomodate for the stochastic nature of ML algorithms and comparing high-dimensional data structures.}

\begin{figure}
  \centering
  \begin{minipage}{0.49\textwidth}
    \centering
    \includegraphics[width=\linewidth]{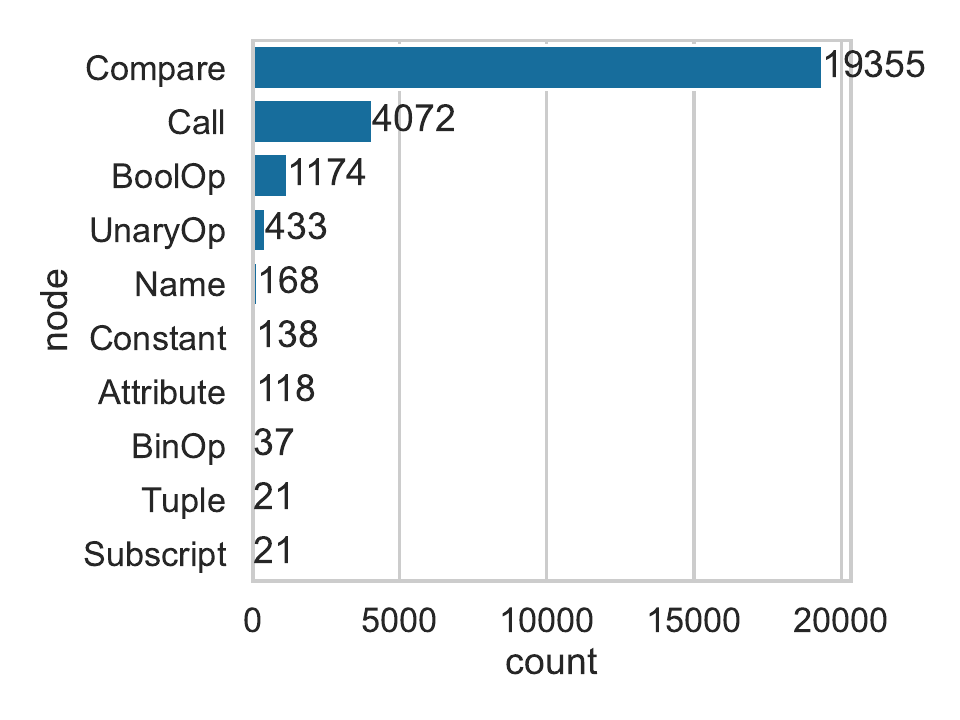}
    \caption{10 most common AST nodes in the test attribute of \lstinline{assert} statements.}
    \label{fig:common-assert-test}
  \end{minipage}
  \hfill
  \begin{minipage}{0.49\textwidth}
    \centering
    \includegraphics[width=\linewidth]{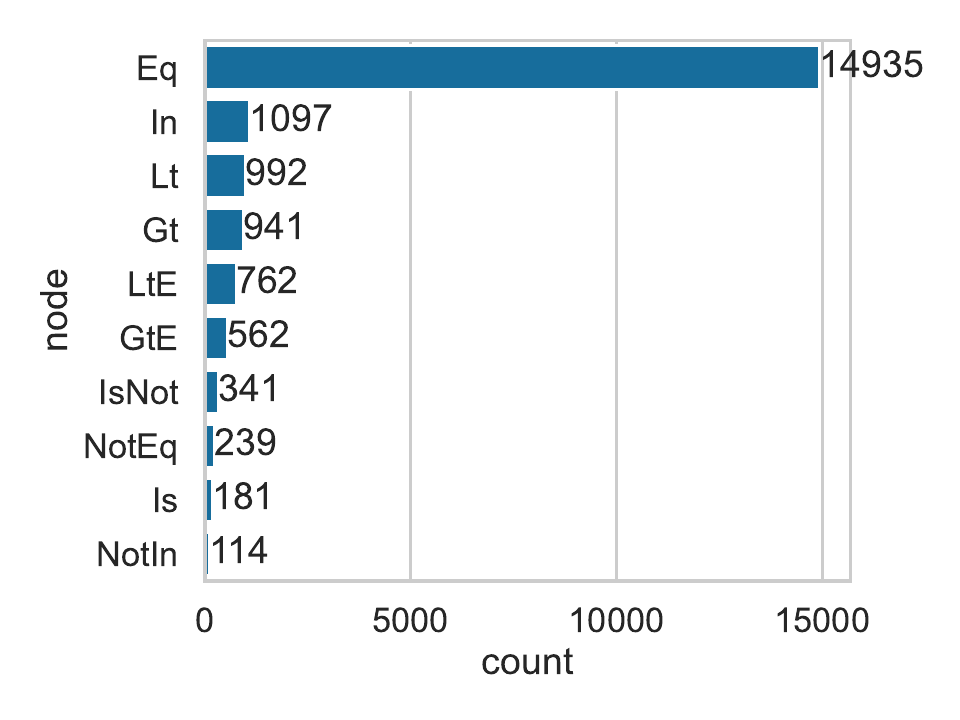}
    \caption{10 most common operators used in \lstinline{assert} statements that perform a comparison check.}
    \label{fig:common-compare-op}
  \end{minipage}
\end{figure}

\begin{table}
  \centering
  \caption{Examples of the most common tests performed in \lstinline{assert} statements.}
  \begin{tabular}{@{}m{0.5\textwidth} m{0.5\textwidth}@{}}
    \toprule
    \emph{\textbf{Comparison Check}}&
    \emph{\textbf{Function Call}}\\
    \midrule

    \lstinline[]$type(train_dataset[0]) == tuple$&
    \lstinline[]$(np.unique(y_test) == [-1, 1]).all()$\\

    \lstinline[]$torch.__version__ > '1.10.0'$&
    \lstinline[]$isinstance(transmat, pd.dataframe)$\\

    \lstinline[]$len(x_dev) == len(y_dev)$&
    \lstinline[]$np.allclose(np.linalg.norm(wio, axis=0), 1)$\\
    \bottomrule
  \end{tabular}
  \label{tab:common-assert-test}
\end{table}

\begin{figure}
  \centering
  \begin{minipage}{0.49\textwidth}
    \centering
    \includegraphics[width=\linewidth]{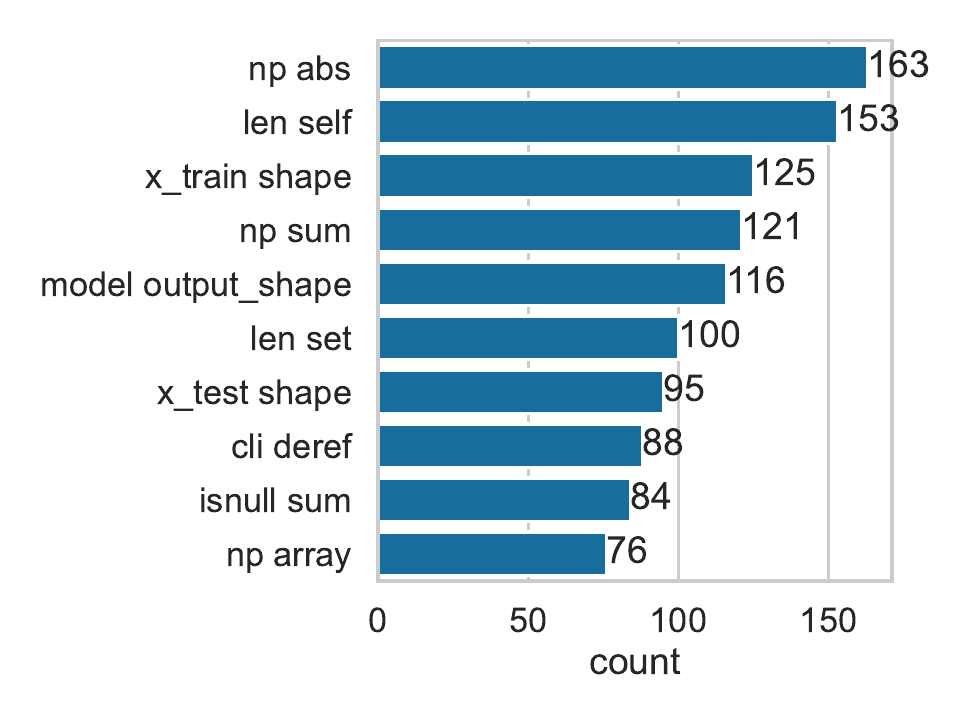}
    \caption{Most common bi-grams that appear in the LHS of comparison statements.}
    \label{fig:common-compare-lhs}
  \end{minipage}
  \hfill
  \begin{minipage}{0.49\textwidth}
    \centering
    \includegraphics[width=\linewidth]{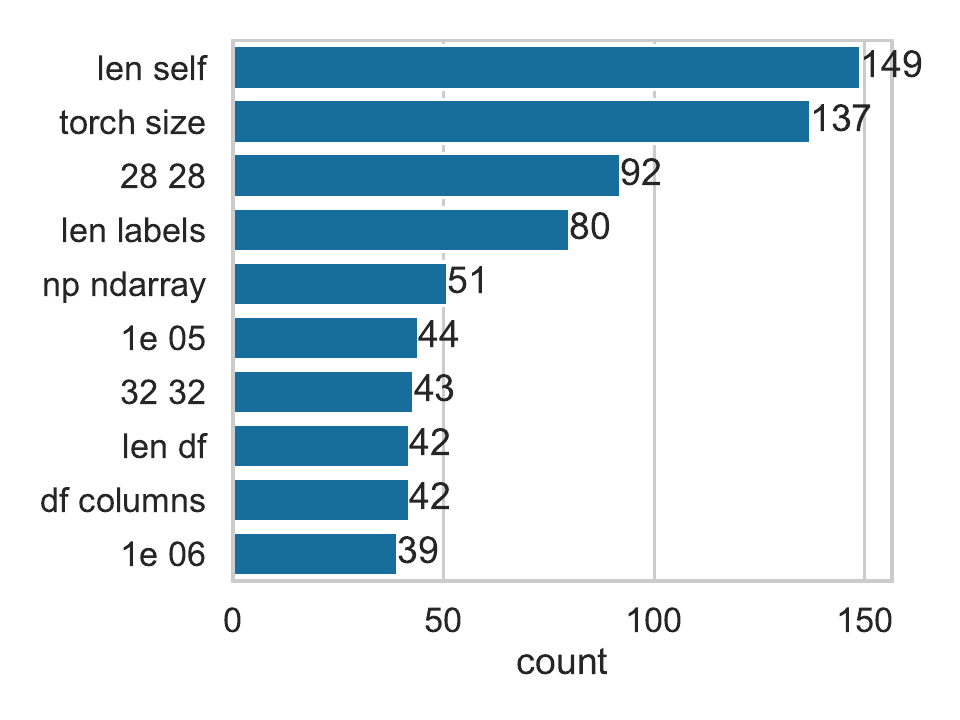}
    \caption{Most common bi-grams that appear in the RHS of comparison statements.}
    \label{fig:common-compare-rhs}
  \end{minipage}
\end{figure}

We focus the remainder of the analysis on the 95\% of the assertions written using the \lstinline{assert} statement. Figure~\ref{fig:common-assert-test} shows the most frequent AST nodes that occur in \lstinline{assert} statements. The figure indicates that the most common test performed in the \lstinline{assert} statements include a comparison check and a function call. Table~\ref{tab:common-assert-test} provides a few examples of assertions for each of the tests above. In Figure~\ref{fig:common-compare-op} we see the most common operators used in assert statements that perform a comparison check. We observe that overwhelming majority of the asserts ensure that two values are equal to each other. Figure~\ref{fig:common-compare-lhs} and Figure~\ref{fig:common-compare-rhs} show the most common bi-grams that appear in the LHS and RHS of the comparison statements respectively. The figures collectively show that the shape and size of data structures are frequently checked.

\begin{figure}
  \centering
  \includegraphics[width=0.5\textwidth]{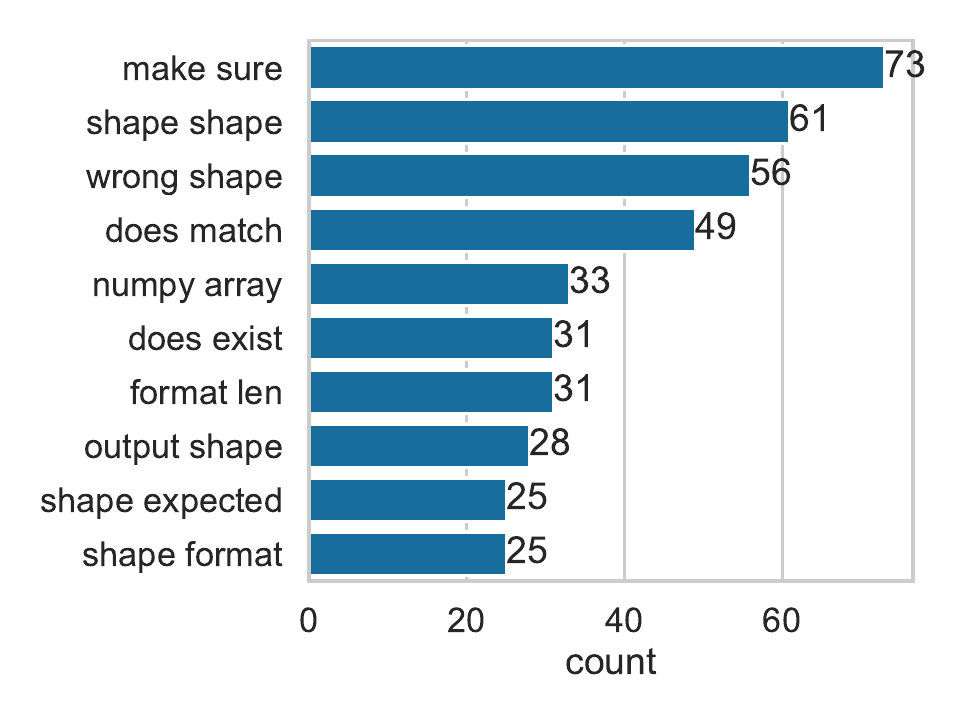}
  \caption{Most common bi-grams in the failure messages of assert statements.}
  \label{fig:common-assert-msgs}
\end{figure}

Of the 95\% assert statements, 28.4 thousand (31.7\%) have a failure message. Figure~\ref{fig:common-assert-msgs} shows the most frequent bi-grams that appear in these failure messages. Similar to the analysis presented above, we removed the duplicate messages prior to analysing the bi-grams resulting in 5.9 thousand unique messages. The bi-grams indicate that the shape and data type of data structures are frequently verified.

\highlight{\textbf{Finding 1.2} 31.7\% of the assertions written using the built-in \lstinline{assert} statement have a failure message. These assertions frequently perform a comparison check or call an external function. The comparison checks frequently verify that the shape and size of data structures is equal to a specified value.}

\begin{figure}
  \centering
  \begin{minipage}{0.49\textwidth}
    \centering
    \includegraphics[width=\textwidth]{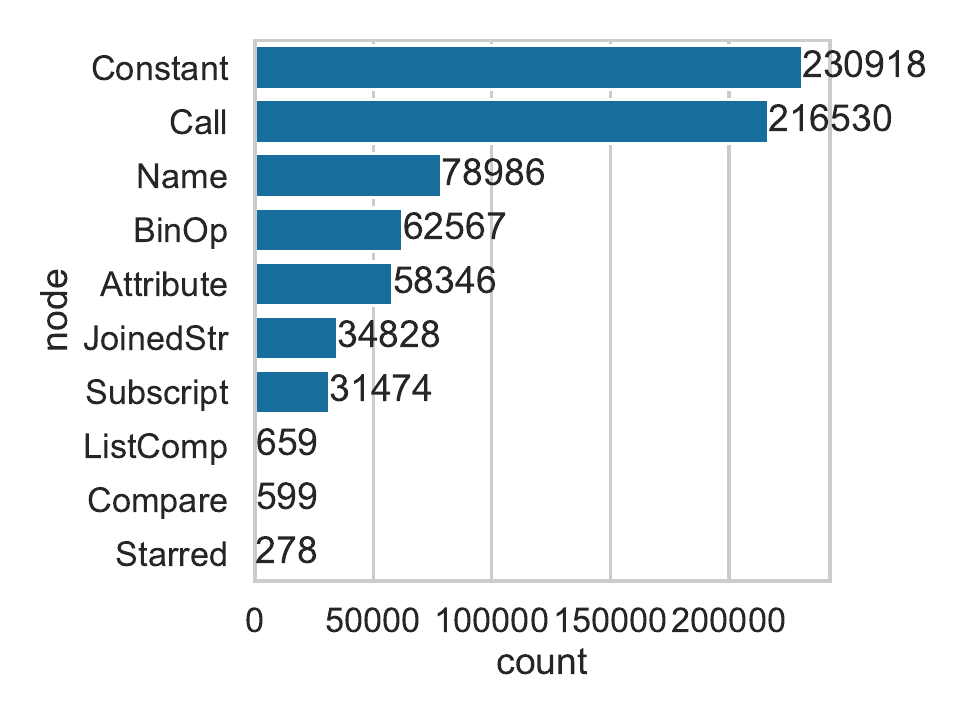}
    \caption{Most common AST nodes inside print statements}
    \label{fig:common-print-nodes}
  \end{minipage}
  \hfill
  \begin{minipage}{0.49\textwidth}
    \centering
    \includegraphics[width=\textwidth]{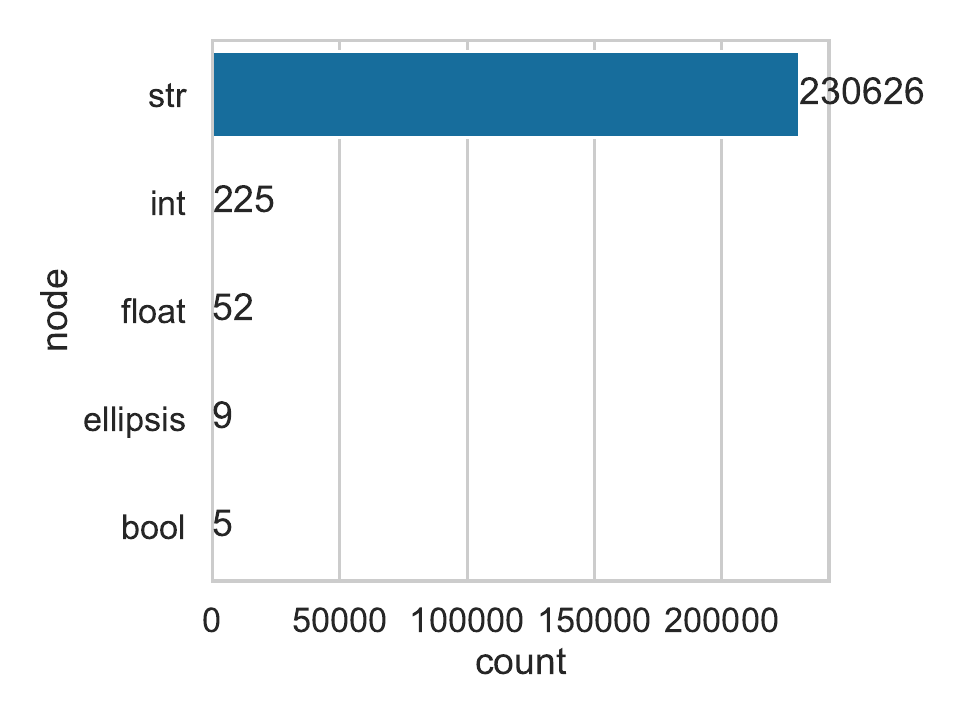}
    \caption{Most common type of constant inside print statements}
    \label{fig:common-print-constant-types}
  \end{minipage}
\end{figure}

\begin{figure}
  \centering
  \begin{minipage}{0.49\textwidth}
    \centering
    \includegraphics[width=\linewidth]{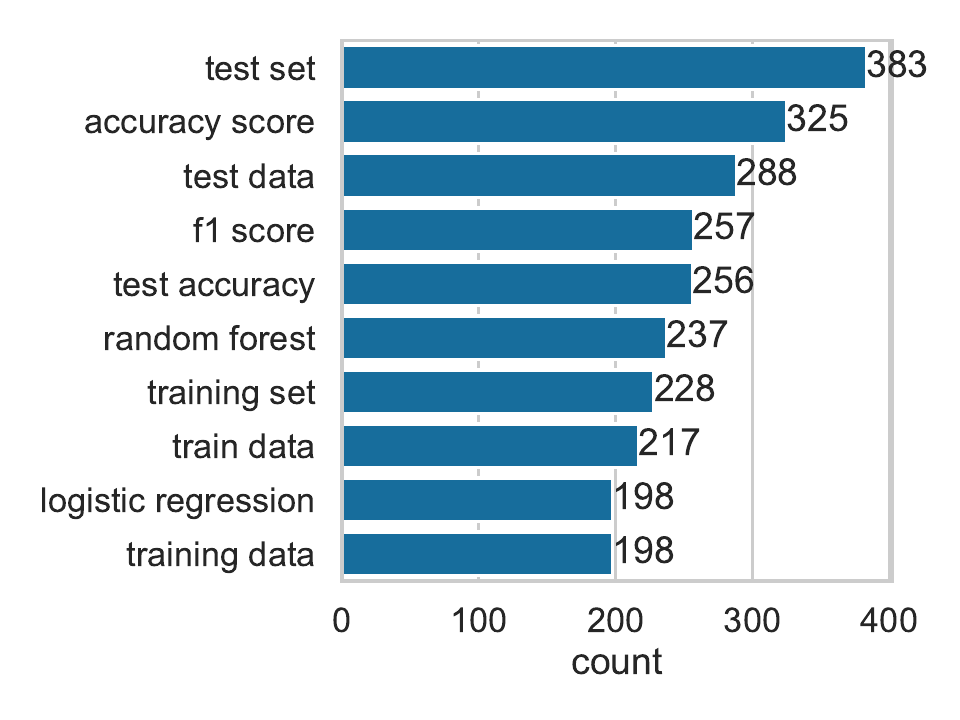}
    \caption{Most common bi-grams that appear in natural text of print statements.}
    \label{fig:common-print-constants}
  \end{minipage}
  \hfill
  \begin{minipage}{0.49\textwidth}
    \centering
    \includegraphics[width=\linewidth]{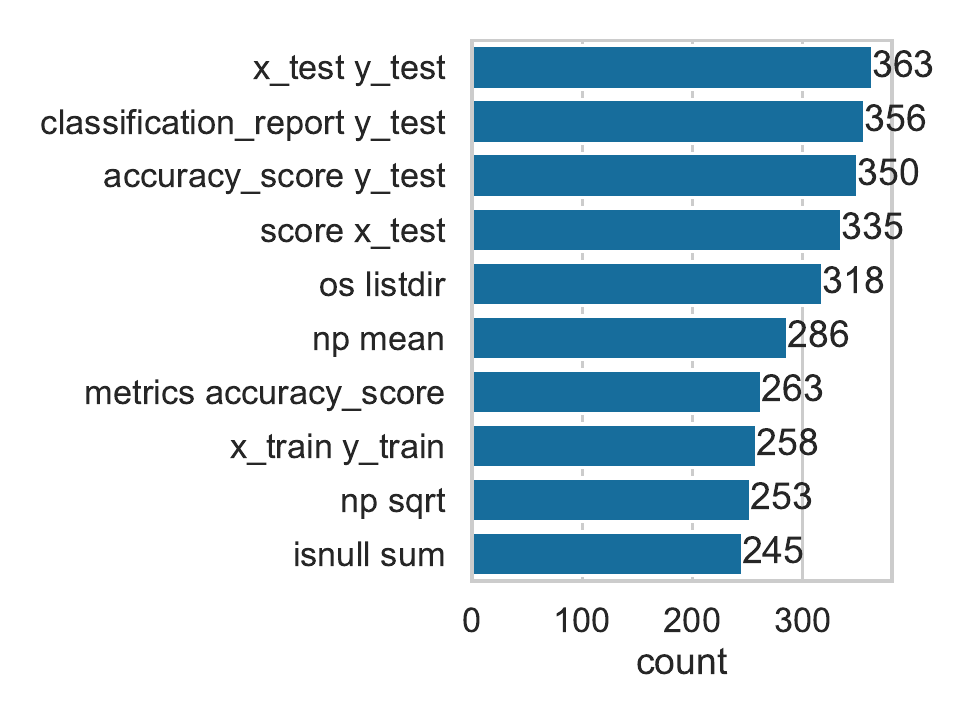}
    \caption{Most common bi-grams that appear in the code of print statements.}
    \label{fig:common-print-not-constants}
  \end{minipage}
\end{figure}

We collected 1.4 million print statements from 180 thousand notebooks (60\% of the total notebooks analysed in this study). Unlike assert statements, print statements do not have a fixed lexical semantic since the Python \lstinline{ast} module interprets them as regular function calls. Additionally, print statements can be written in several forms---the \lstinline{print} function may have multiple arguments, any form of string interpolation supported by Python or output from function calls. Figure~\ref{fig:common-print-nodes} shows that the most common AST nodes inside of print statements are constants followed by function calls. Figure~\ref{fig:common-print-constant-types} further validates that the constants are almost entirely of type string. Based on these observations, we dissect the print statements into two components---text written in natural language (from the \emph{Constant} nodes) and code (from all other nodes). Figure~\ref{fig:common-print-constants} and Figure~\ref{fig:common-print-not-constants} present the most common bi-grams that appear in the natural text and code of print statements respectively. Both figures collectively indicate that print statements are frequently used to print various performance metrics of ML models.

\highlight{\textbf{Finding 1.3} Print statements are most commonly used to display various performance metrics of ML models.}

\begin{figure}
  \centering
  \includegraphics[width=0.5\textwidth]{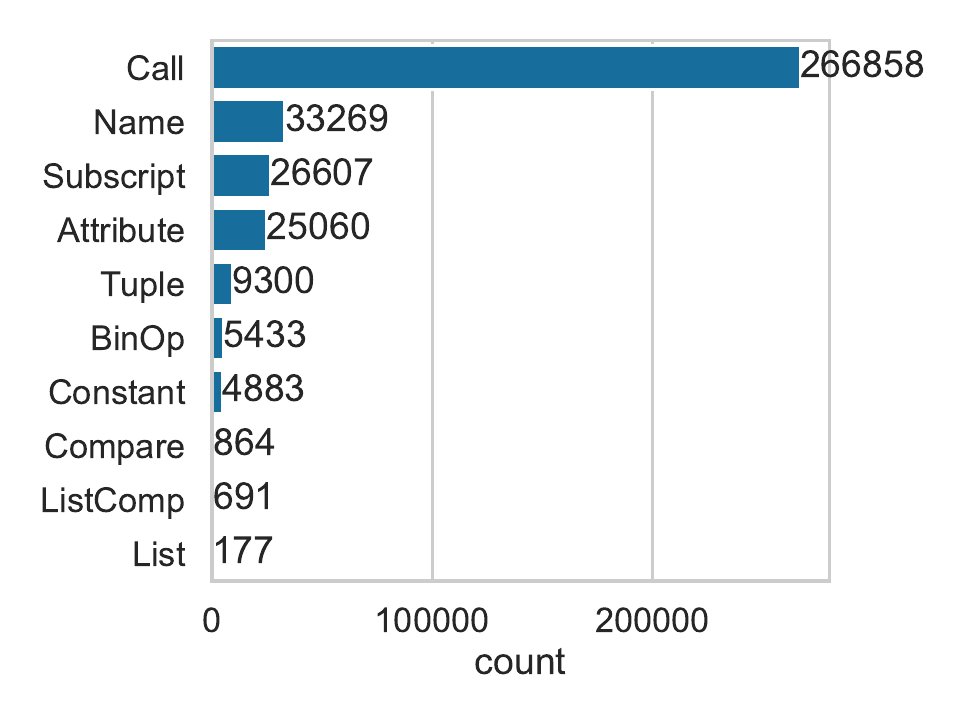}
  \caption{Most common AST nodes in the last cell statements.}
  \label{fig:common-last-nodes}
\end{figure}

\begin{figure}
  \centering
  \begin{minipage}{0.49\textwidth}
    \centering
    \includegraphics[width=\linewidth]{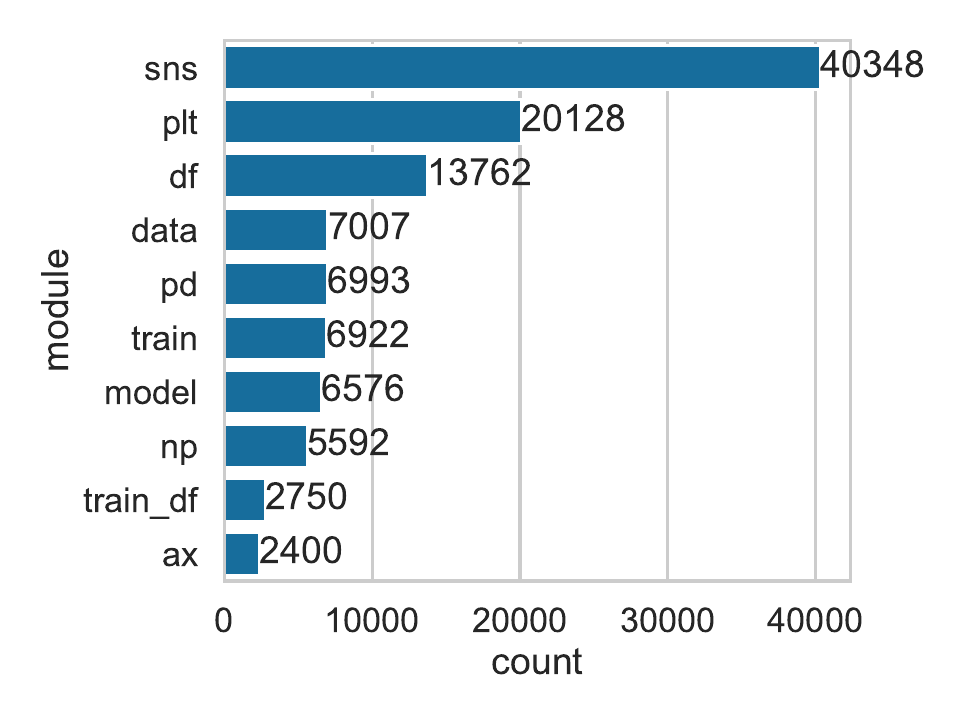}
    \caption{Most common Python modules and objects used in the \emph{Call} nodes of last cell statements.}
    \label{fig:common-last-modules}
  \end{minipage}
  \hfill
  \begin{minipage}{0.49\textwidth}
    \centering
    \includegraphics[width=\linewidth]{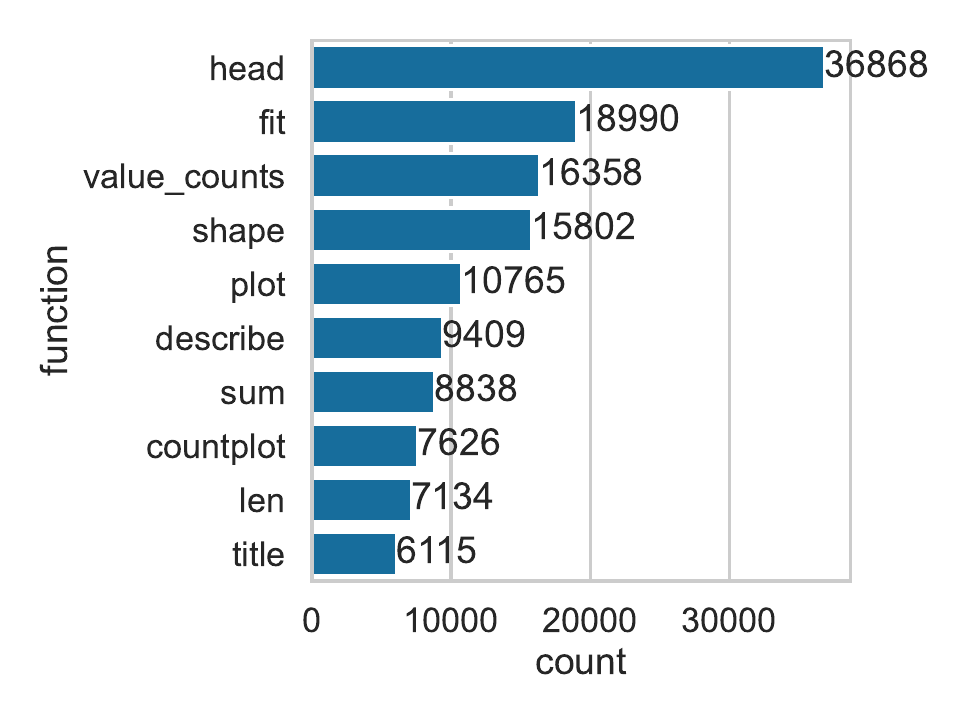}
    \caption{Most common functions used in the \emph{Call} nodes of last cell statements.}
    \label{fig:common-last-functions}
  \end{minipage}
\end{figure}

We collected 1 million last cell statements from 119 thousand notebooks (40\% of the total notebooks analysed in this study). Figure~\ref{fig:common-last-nodes} shows that \emph{Call} nodes frequently occur in the AST of last cell statements. Figure~\ref{fig:common-last-modules} shows the most common Python modules and objects while Figure~\ref{fig:common-last-functions} show the most common functions used inside these \emph{Call} nodes. The figures highlight the prominence of data visualization libraries seaborn (\lstinline{sns}) and matplotlib (\lstinline{plt}) and visualization functions such as \lstinline{plot} and \lstinline{countplot} in the notebooks. The data analysis tool pandas (represented by \lstinline{df} and \lstinline{pd}) also features prominently with many data exploration and preprocessing functions such as \lstinline{head}, \lstinline{value_counts} and \lstinline{describe} being used frequently. The presence of \lstinline{train} and \lstinline{model} Python objects along with the \lstinline{fit} function indicates the use of ML libraries.

\highlight{\textbf{Finding 1.4} Libraries and functions for data visualization, data exploration and machine learning are most common in last cell statements.}

\begin{figure}
  \centering
  \begin{minipage}{0.49\textwidth}
    \centering
    \includegraphics[width=\linewidth]{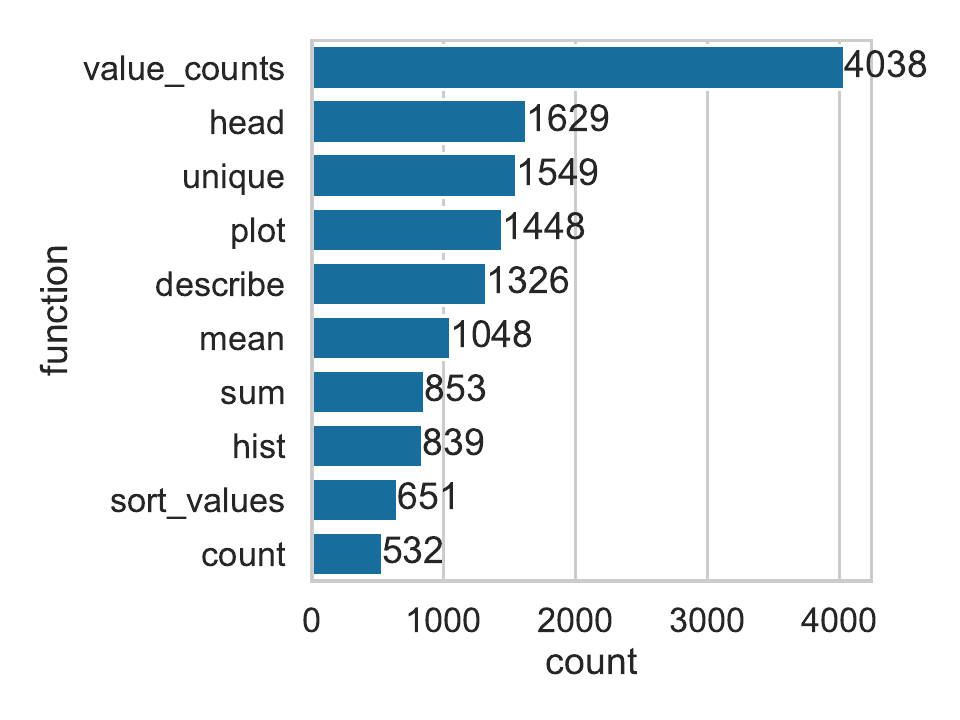}
    \caption{Most common functions called on pandas dataframes.}
    \label{fig:common-last-df-functions}
  \end{minipage}
  \hfill
  \begin{minipage}{0.49\textwidth}
    \centering
    \includegraphics[width=\linewidth]{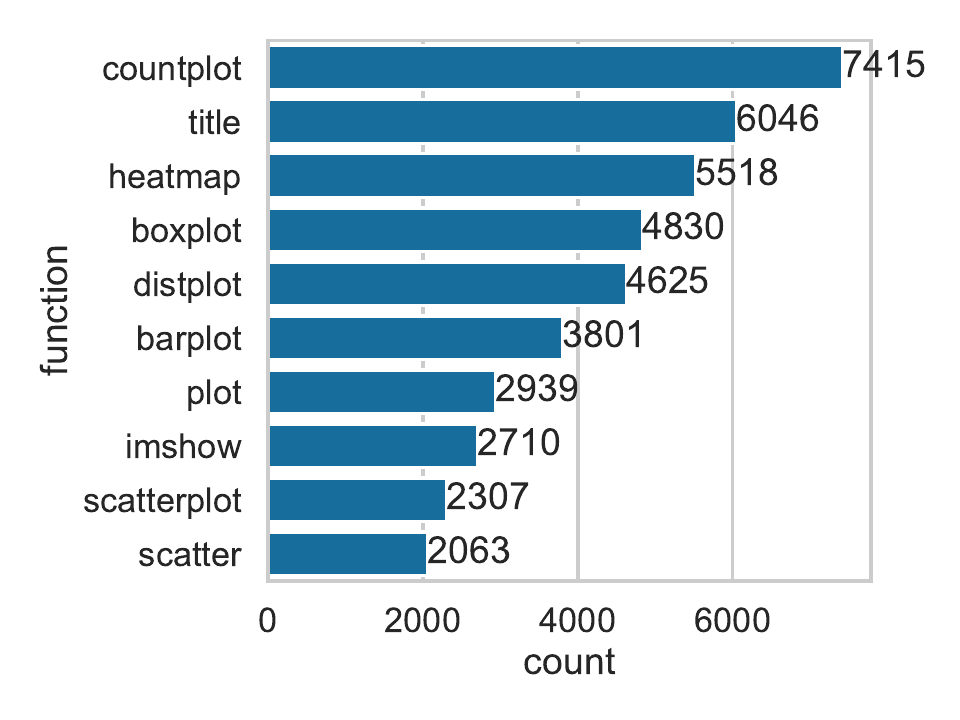}
    \caption{Most common visualization functions used in last statements.}
    \label{fig:common-last-visualization-functions}
  \end{minipage}
\end{figure}

We further analysed the most common functions called on pandas dataframes by isolating the last cell statements that contain \lstinline{data} or \lstinline{df} objects. This is shown in Figure~\ref{fig:common-last-df-functions} which reveals a focus on exploratory data analysis and statistical operations. The frequent use of functions such as \lstinline{value_counts}, \lstinline{head}, and \lstinline{unique}, indicate an emphasis on understanding the structure and content of the data. Visualization functions like \lstinline{plot} and \lstinline{hist} suggest that practitioners often create quick visualizations directly from dataframes. Descriptive statistics are also prominent, with functions like \lstinline{describe}, \lstinline{mean}, and \lstinline{sum} being commonly used.

Similarly, we also analysed the most common visualization functions used in notebooks by isolating the last cell statements where the module is \lstinline{sns} or \lstinline{plt}. This is shown in Figure~\ref{fig:common-last-visualization-functions} which indicates that categorical data visualization appears to be a primary focus, with \lstinline{countplot} and \lstinline{barplot} ranking high among the most frequently used functions. The prominence of \lstinline{heatmap} and \lstinline{boxplot} functions indicates a regular need for visualizing correlations and statistical distributions. For continuous data, functions like \lstinline{distplot}, \lstinline{scatterplot}, and \lstinline{scatter} are commonly used. The presence of \lstinline{imshow} suggests that image-related visualizations are also significant in many notebooks.

\highlight{\textbf{Finding 1.5} Common functions called on pandas dataframes focus on exploratory data analysis, statistical operations and quick visualizations. In visualizations, functions for categorical data, correlation and distribution are frequently used.}

\subsection{(RQ2) How is explicit feedback from assert statements used to validate ML code written in Jupyter notebooks?}~\label{sec:result-explicit}

\subsubsection{Data shape check ($N = 26$)}

Data shape checks can be considered the ``Swiss army knife'' of assertions, as they are ubiquitous and serve multiple purposes throughout the MLDL. We encountered a vast array of assert statements that verify the dimensions of various data structures, including input features (\emph{A17, A90}), labels (\emph{A5, A29}), predictions, images (\emph{A76, A93}), sequences, and embeddings (\emph{A76, A93}). These assertions ensure that the data adheres to the expected format and dimensions required by different components of the machine learning pipeline.

\highlight{\textbf{Finding 2.1} 26 assertions verify the dimensions of various data structures such as input features, labels, predictions, images, sequences, and text embeddings. These checks ensure data consistency and compatibility throughout the ML pipeline, preventing runtime errors and ensuring accurate model training and evaluation.}

\subsubsection{Data validation check ($N = 14$)}

We found assert statements that perform various validation checks on data structures, such as NumPy arrays and Pandas dataframes. Common data validation checks include verifying the presence of specific values or ranges within arrays or columns(\emph{A41, A65}) and verifying the equality or closeness of arrays to a target value within a specified absolute tolerance (\emph{A44, A46}). Some assertions also focus on validating the uniqueness or cardinality of values (\emph{A52}) while others validate the presence of specific values or conditions within data structures (\emph{A73}).

\highlight{\textbf{Finding 2.2} Fourteen assertions perform validation checks on data structures, ensuring the presence of specific values, verifying value ranges, and confirming uniqueness or cardinality of values. These checks catch potential errors or inconsistencies early, ensuring data meets specific criteria or constraints for robust ML applications.}

\subsubsection{Model performance check ($N = 11$)}\label{sec:assert-model-perf}

\begin{lstlisting}[caption={Assertion \emph{A58} used to check that the accuracy of a model is close to the specified value. The use of \lstinline{np.isclose} allows for small deviations in the accuracy thus accounting for the stochastic nature of ML models.}, label={lst:A58}]
assert np.isclose(accuracy, 0.9666666666666667)
\end{lstlisting}

This study finds several assertions used to test model performance against predefined thresholds for key metrics such as accuracy, precision, recall and F1 score. Range checks confirm that the performance metrics fall within acceptable boundaries, indicating good predictive performance without overfitting (\emph{A38}). Direct comparisons of model outputs with expected results, validate the model's predictive accuracy and reliability under operational conditions (\emph{A15}). Other assertions ensure that the learned parameters align closely with theoretically or empirically derived values, further cementing the model’s statistical validity (\emph{A19}). Finally, assertions are also used to check for the expected number of neighbours as an indirect measure of model performance in specific scenarios like clustering (\emph{A7}).

The use of \lstinline{np.isclose} in Listing~\ref{lst:A58} is particularly noteworthy. This method accounts for the stochastic nature of many ML models, where slight variations in performance metrics can occur due to differences in initial conditions, random seed settings, or inherent randomness in algorithms. By allowing a small tolerance in the comparison, \lstinline{np.isclose} ensures that the model's performance is consistently close to the expected benchmark despite the stochastic elements.

\highlight{\textbf{Finding 2.3} Eleven assertions evaluate model performance against predefined thresholds for accuracy, precision, recall, F1 score, and other metrics. These checks validate the model's predictive accuracy and reliability, ensuring it meets expected performance standards despite the stochastic nature of ML models.}

\subsubsection{Existence check ($N = 8$)}

Existence checks are primarily used during the data preprocessing stage in the MLDL to ensure data integrity before analysis and modelling. These checks primarily focus on verifying the presence of necessary columns and the absence of missing values within those columns (\emph{A23, A42, A79, A86}). Such validations are especially crucial after data preprocessing steps, where transformations might inadvertently introduce \lstinline{NaN} values (\emph{A50, A51, A63}) or result in an empty dataframe (\emph{A43}). These checks are also crucial for maintaining the accuracy and efficiency of data handling processes, ensuring that the datasets are ready for robust ML applications.

\highlight{\textbf{Finding 2.4} Eight assertions ensure the presence of specific columns in the dataset and the absence of missing values. These checks prevent operations on empty datasets and confirm the absence of \texttt{NaN} values, maintaining data integrity and reliability for model training.}

\subsubsection{Resource check ($N = 7$)}~\label{sec:result-rq2-resource-check}

\begin{lstlisting}[caption={Assertion \emph{A37} used to ensure that an ML model has not reached an inconsistent state due to out-of-order or re-execution of code cells.}, label={lst:A37}]
assert svm.fit_status_ == 0, 'Forgot to train the SVM!'
\end{lstlisting}

Resource check assertions are used to ensure the availability and validity of essential resources, preventing runtime errors, and maintaining the integrity of the data, models, and visualizations. One set of assertions verify the existence of files on the file system, such as pre-trained models (\emph{A10}) or data files (\emph{A74}). Subsequently, assertions are employed to validate the successful loading of models (\emph{A14}) and the completion of model training processes (\emph{A37}).

A unique aspect of working with notebooks is the ability to re-run cells while experimenting, which can lead to unintended consequences. Assertion \emph{A37} presented in Listing~\ref{lst:A37} addresses this scenario by checking if the SVM model has been properly trained before proceeding with further operations. This check prevents inconsistent model states from executing cells out of order or multiple times. Other interesting assertions unique to computational notebooks include verifying the presence of data in visualizations (\emph{A60}) and explicitly verifying the versions of critical libraries (\emph{A18, A67}).

\highlight{\textbf{Finding 2.5} Seven assertions confirm the availability of essential resources like pre-trained models, data files, successful model loading, and completion of training processes. These checks prevent runtime errors and ensure proper code execution, maintaining the reliability and robustness of the ML workflows.}

\subsubsection{Type check ($N = 5$)}

Type check assertions are used to maintain data integrity, ensure compatibility between different components of the ML pipeline, and preventing errors and unexpected behaviour. Type checks are commonly performed to ensure that the input features are PyTorch float tensors or NumPy arrays, a requirement for many deep learning models and scientific computing operations respectively (\emph{A2, A88}). Type checks are also applied to verify the type of the ML models (\emph{A40}), data structures and intermediate objects (\emph{A35}) and individual columns or elements within a dataset (\emph{A81}).

\highlight{\textbf{Finding 2.6} Five assertions validate the data types of input features, models, and intermediate objects. These checks ensure compatibility with ML models and operations, preventing errors from incompatible methods or attributes, thus maintaining the reliability of the ML workflow.}

\subsubsection{Mathematical property check ($N = 4$)}

When working with neural networks and other statistical models, it is imperative to ensure that mathematical properties are preserved throughout operations involving arrays and matrices. This rigour is captured though assertions that ensure convolution operations are dimensionally consistent (\emph{A3}), checking for the symmetry of matrices (\emph{A64}) and monitoring the standard deviation of outputs to decide the appropriateness of using batch normalization (\emph{A25}). These assertions are not just precautionary but are vital for confirming that the underpinnings of ML algorithms align with expected mathematical principles.

\highlight{\textbf{Finding 2.7} Four assertions validate key mathematical properties of data and model outputs. These include ensuring dimensional consistency in convolution operations, monitoring standard deviation for batch normalization suitability, verifying uniformity in computations, and checking matrix symmetry. These checks confirm the robustness and reliability of model computations.}

\subsubsection{Batch size check ($N = 3$)}

In the context of neural networks, the practice of using batch processing is a key strategy to enhance computational efficiency and hardware utilization. Batching stabilizes the learning curve by updating the model weights based on the average gradient of a batch thereby optimizing learning and enhancing generalizability of the model. Ensuring the batch size matches the hardware capacity is crucial to avoid memory overflow issues that can halt training. Assertions are used to check that the batch size divides evenly into the training dataset size which ensures that each training epoch is computationally efficient (\emph{A21}). Additionally, assertions are used to ensure that the size of the training data is larger than the batch size (\emph{A70}) and that image dimensions are suitably divisible by the patch size which is essential for certain convolutional networks (\emph{A28}).

\highlight{\textbf{Finding 2.8} This study identifies three assertions validating batch size configurations. These assertions ensure batch sizes are divisible by dataset sizes, image dimensions are appropriately divided by patch sizes, and the number of images exceeds the batch size. These checks are crucial for optimizing computational efficiency and ensuring stable and smooth model training.}

\subsubsection{Network architecture check ($N = 3$)}

When integrating pre-trained models or leveraging transfer learning techniques, it is crucial to ensure the compatibility of the custom neural network architecture with the pre-trained components. This is particularly important when combining convolutional layers from different sources, as the spatial dimensions and channel configurations must align correctly. Checks include assertions that verify the output and input dimensions of consecutive convolutional layers match (\emph{A11}), checking that shape of the activations have the right number of channels (\emph{A62}) and that the appropriate regularization parameter is set (\emph{A75}). Such architectural checks are crucial for preventing potential issues during the forward pass of the neural network and ensuring that the custom and pre-trained components are compatible with each other.

\highlight{\textbf{Finding 2.9} Three assertions verify the consistency of neural network architecture. These include ensuring weight dimensions match across layers and that activation shapes and regularization parameters are as expected. These checks ensure that the neural network is correctly configured and integrated with pre-trained models, maintaining the integrity of data flow during training.}

\subsubsection{Data leakage check ($N = 1$)}

\begin{lstlisting}[caption={Assertion \emph{A33} used to ensure that the training and validation sets do not contain any overlapping values.}, label={lst:A33}]
assert len(
  set(tr_df.PetID.unique()).intersection(valid_df.PetID.unique())
) == 0
\end{lstlisting}

Ensuring the absence of data leakage between training and validation datasets is a fundamental best practice in ML to prevent overfitting. To guarantee that the model can generalize effectively to new examples, it is crucial to verify that the training and validation sets are completely distinct with no shared examples. By strictly separating these datasets, the validation phase provides an unbiased evaluation of the model's ability to generalize to data similar to, but not identical to that which it was trained on. This is demonstrated by Assertion \emph{A33} in Listing~\ref{lst:A33} which ensures that there are no overlapping \lstinline{PetID} values between the training set and the validation set, thereby preventing any possibility of data leakage.

\highlight{\textbf{Finding 2.10} This study identifies one assertion to ensure there are no overlapping values between training and validation sets to prevent data leakage. Preventing data leakage is essential to avoid overfitting and ensure the model's ability to generalize to new data. This practice maintains the integrity and validity of the model evaluation metrics.}

\subsection{(RQ3) How is implicit feedback from print statements and last cell statements used when writing ML code in Jupyter notebooks?}~\label{sec:result-implicit}

\subsubsection{Model performance check ($N = 33$)}

During the model development phase, the performance of an ML model is often printed to facilitate comparisons against other models or variations. Throughout this continuous experimentation phase, authors may adjust model parameters or modify the data by engineering new features. The model is then re-trained to check if these changes lead to performance improvements. Besides the \emph{accuracy} of the model (\emph{P3, P18}), we also see checks for the \emph{Root Mean Square Error (RMSE)} (\emph{P6}) and the use of the classification report (\emph{P50}) provided by scikit-learn\footnote{https://scikit-learn.org/stable/modules/generated/sklearn.metrics.classification\_report.html} which reports the \emph{accuracy}, \emph{precision}, \emph{recall} and \emph{f1 score} for all labels present in the target feature. In addition, we find the use of heatmaps (\emph{L3}) to visualize the confusion matrix of a multi-label classification task and custom functions (\emph{L52}) to manually evaluate an image classification model on a random sample of input images.

\highlight{\textbf{Finding 3.1} The study identifies 33 implicit checks to monitor the performance of trained ML models. These include outputs that print key performance metrics, a normalized heatmap used to visualize the confusion matrix of a model's predictions and a custom function written for manual evaluation of model predictions on random input samples. These checks are used throughout the iterative model development phase, enabling practitioners to compare different models or variations, adjust parameters, and refine features to improve performance.}

\subsubsection{Data distribution check ($N = 7$)}~\label{sec:data-distribution-output}

Understanding the distribution of data is crucial for making informed decisions about necessary transformation steps in data analysis. For instance, visualizations can efficiently determine if scaling, normalizing, or handling outliers is needed. During the exploratory data analysis phase, visualizations and pandas dataframes are commonly used to assess the distribution of specific columns (\emph{L9}), helping identify features related to the target variable (\emph{L2, L14})and informing feature inclusion in model training. Additionally, descriptive statistics are used post-transformation to manually verify the effectiveness of data transformation steps and ensuring that the data conforms to the expected format for further analysis (\emph{L48, L25}).

\highlight{\textbf{Finding 3.2} The study identifies seven implicit data distribution checks, including visualizations like categorical plots, kernel density estimate plots and count plots, and statistical summaries from pandas dataframes. These checks are used for making informed decisions about data transformations such as scaling, normalizing and handling outliers, ensuring data integrity and improving model training reliability during the exploratory data analysis phase.}

\subsubsection{Resource check ($N = 7$)}\label{sec:implicit-resource-check}

Similar to the assertions identified in Section~\ref{sec:result-rq2-resource-check}, we also find print and last cell statements that validate the availability of essential resources on the system where the notebooks are being executed. This includes checks for the availability of compute resources such as a GPU or a TPU (\emph{P68, P82, P86}), verifying that a dataset or a pre-trained model has been successfully loaded into memory (\emph{P107, L66}) and ensuring that a certain version of an external library is present on the system (\emph{P71}).

\highlight{\textbf{Finding 3.3} The study identifies seven implicit checks used to verify resource availability on the systems where the Jupyter notebooks are being executed. These checks include confirming the availability of compute resources like GPUs and TPUs, ensuring datasets or pre-trained models are successfully loaded, and verifying the presence of specific library versions. These checks are used for ensuring that the computational environment is correctly set up, preventing execution errors, and facilitating smooth workflow execution.}

\subsubsection{Spot check ($N = 5$)}

Performing value or spot checks can be an essential practice for ensuring data integrity and model accuracy at various stages of development. These checks are crucial for verifying that operations such as data transformations, model predictions, and feature engineering are functioning as expected. Common checks include verifying the number of features in the data matches expectations after applying feature selection or dimensionality reduction techniques (\emph{L60}), assessing the correctness of one-hot encoding (\emph{P114}) and ensuring that data retains its expected properties after manipulation (\emph{P64}).

\highlight{\textbf{Finding 3.4} This study finds five implicit spot checks performed at various stages of the MLDL. Spot checks are used to ensure data integrity by verifying data transformation and feature engineering steps meet expectations.}

\subsubsection{Model training check ($N = 4$)}

During training, it is common practice to monitor the progress periodically and use this feedback to adjustment training parameters or stop training early to prevent overfitting. This can be done by checking the best parameters found through tuning methods (\emph{L42}), fitting the model to the training data (\emph{L31}) or printing the training loss and accuracy to give real-time feedback on the learning efficacy of the model (\emph{L8}).

\highlight{\textbf{Finding 3.5} The study identifies four implicit checks used to monitor model training progress. These checks provide real-time feedback on training loss and accuracy, validate model adherence to expected data patterns, and identify optimal parameters through tuning methods. These checks facilitate timely adjustments during the training process to prevent overfitting and ensure model performance optimization.}

\subsubsection{Missing value check ($N = 3$)}

Checking for missing values is a critical step in the preprocessing phase of an ML project because it significantly influences the quality and performance of the model. Missing data can lead to biased or incorrect conclusions if not handled properly, potentially skewing the model's performance by training on incomplete or non-representative samples~\citep{shome2022data}. Many ML algorithms demand complete numerical datasets to perform calculations such as matrix multiplication. Missing values interrupt these calculations, leading to errors or the inability to execute algorithms entirely. These checks can be performed in code (\emph{P74, L36}) and using visualizations such as a heatmap (\emph{L12}).

\highlight{\textbf{Finding 3.6} The study identifies three implicit checks for missing data. Missing values can lead to biased or incorrect conclusions as many ML algorithms require complete numerical datasets for accurate calculations. Identifying and addressing missing data during preprocessing is essential to avoid errors and improve model reliability.}

\subsubsection{Shape check ($N = 3$)}

Ensuring that data dimensions align with expectations is important particularly following data pre-processing or transformation steps. Primarily, the number of features in the training set must match those in the testing set (\emph{P4, P32, P117}). This alignment is crucial because statistical ML models are trained on specific data dimensions and expect the same dimensional structure during inference to perform accurately. Similarly, in neural network architectures, the configuration of input layers depends directly on the shape of the training data, dictating the number of input neurons needed. Furthermore, the correspondence between the number of test examples and their respective labels is essential for accurately computing performance metrics.

\highlight{\textbf{Finding 3.7} The study identifies three implicit checks used to validate the shape of data. These checks are used to ensure data dimensions align with expectations, particularly after preprocessing or transformation steps.}

\subsubsection{Data relationship check ($N = 2$)}~\label{sec:linear-relation-output}

\begin{table}
  \centering
  \caption{Last cell statements used to verify linear relationship between features in the data.}
  \begin{tabular}{@{}m{0.05\textwidth} m{0.4\textwidth} m{0.4\textwidth}@{}}
    \toprule
    \emph{\textbf{Key}}&
    \emph{\textbf{Code}}&
    \emph{\textbf{Output}}\\
    \midrule

    L6&
    \lstinline[]$b = sns.relplot(x='SIZE', y='Cash', hue='CLARITY', alpha=0.9, palette='muted', height=8, data=raw_data)$&
    \includegraphics[width=\linewidth]{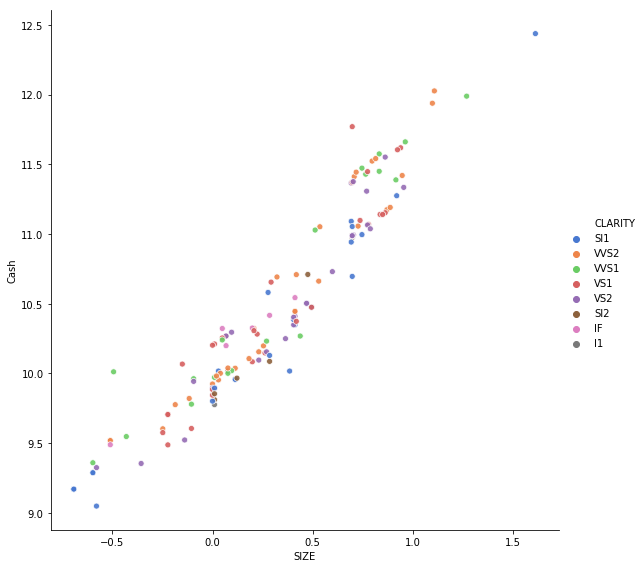}\\

    L10&
    \lstinline[]$sns.regplot(x='X4 number of convenience stores', y='Y house price of unit area', data=data)$&
    \includegraphics[width=\linewidth]{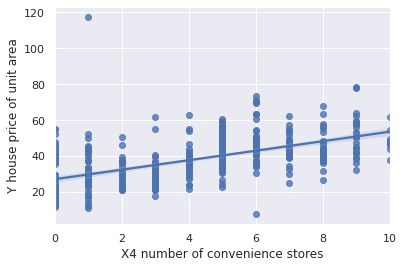}\\
    \bottomrule
  \end{tabular}
\label{tab:linear-relation-check}
\end{table}

Linear ML models achieve optimal performance when the target variable can be expressed as a linear combination of the input features. However, features within a dataset that exhibit a linear relationship are considered redundant as they convey the same information to the model during training. Consequently, the feature engineering stage often involves removing such features to create a more efficient training dataset~\citep{shome2022data}. Table~\ref{tab:linear-relation-check} illustrates two last cell statements found in this study for identifying linear relationships among features in the data. In addition to the code, the raw output of the cell is also shown since both case-studies produce a visualization.  In \emph{L6}, the practitioner assesses the linearity between the features \texttt{Cash} and \texttt{SIZE}. \emph{L10} is a visualization of the feature \texttt{X4} alongside the target variable \texttt{Y}, accompanied by a fitted linear regression model.

\highlight{\textbf{Finding 3.8} The study identifies two last cell statements used to verify linear relationships between features in the data. Visualizations such as a scatter plot with a regression line and a line plot are employed to assess the linearity between features. These checks are used for identifying and potentially removing redundant features or performing feature selection to ensure that the training dataset is optimized for linear ML models.}

\subsubsection{Type check ($N = 2$)}

Type checks are crucial not only for confirming the current state of the data but also for determining the specific transformations required to make the data suitable for subsequent modelling steps. Ensuring data types align with model requirements is a fundamental step in the exploratory data analysis phase. Since all ML algorithms perform mathematical operations, they require the input data to be entirely numerical to avoid computational errors (\emph{P43, L71}). The process of checking data types helps in identifying the necessary pre-processing or transformation steps to convert the data into a form that is compatible with these models. For example, numerical features often require normalization to bring them within a specific scale and categorical features must be converted into a numerical format through methods like one-hot encoding to ensure they can be effectively integrated into the model.

\highlight{\textbf{Finding 3.9} The study identifies two implicit checks used to verify the data types of features in the dataset. These checks are used to determine the necessary preprocessing or transformation steps required to make the data suitable for training. Type checks are also used to ensure that all features are numerical prior to performing mathematical operations such as matrix multiplication, which is used in several ML models.}

\subsubsection{Execution time check ($N = 1$)}
\begin{lstlisting}[caption={Print statement \emph{P66} used to check the total execution time of training an ML model with various hyper-parameter configurations.}, label={lst:P66}]
print('Total Run Time:')
\end{lstlisting}

While training multiple ML models for a single task is common, choosing the best model considers not only its performance but also its training speed. Faster training times are more cost-effective for iterative tasks and promote sustainability in the long run~\citep{shome2022data}. Listing~\ref{lst:P66} presents print statement \emph{P66} that prints the total execution time of a code cell which trains an ML model with various hyperparameter configurations using a grid-search.

\highlight{\textbf{Finding 3.10} The study identifies one print statement used to check the total execution time of training and cross-validating an ML model. The print statement can be used to evaluate the efficiency of model training, especially when multiple models are trained iteratively.}

\subsubsection{Network architecture check ($N = 1$)}

Printing the structure of a neural network (\emph{P92}) serves crucial purposes such as verifying the correct implementation of the architecture, aiding in debugging by ensuring layer compatibility, and providing a clear, human-readable format for documentation and educational insights. It helps in verifying configurations like layer dimensions and operations before initiating computationally intensive training and ensuring that the model is set up efficiently for the intended tasks.

\highlight{\textbf{Finding 3.11} The study identifies one print statement used to show the architecture of a neural network. Such outputs can be used during the model development phase to ensure compatibility between the layers of a neural network, for understanding and learning about the architecture of pre-trained models and for documentation purposes.}

\section{Discussion}\label{sec:discuss}

Drawing from the results of this study, we discuss implications and future work in this section.

\subsection{Documentation and Technical Debt}

Jupyter notebooks are primarily used to explore different ideas or approaches, often neglecting software development best practices in favour of rapid experimentation~\citep{kery2018story,rule2018exploration,pimentel2019large-scale}. Once the code is finalized, it is typically transferred into a script that is run in a production environment. Alternatively, the notebook may contain significant exploratory work such as developing a model or performing data exploration. To effectively communicate the information present in the notebooks with clients, collaborators or teammates, the practitioners add annotations and structure to the notebook using markdown cells~\citep{kery2018story,rule2018exploration}. In both scenarios, practitioners need to revisit the notebooks to obtain up-to-date and accurate requirements, their rationale and design decisions, and tests---artefacts which are usually missing in the original notebooks~\citep{pimentel2019large-scale,psallidas2019data,grotov2022large-scale}.

Our findings indicate that structured information can be effectively derived from the feedback mechanisms available in Jupyter notebooks. The results demonstrate that practitioners utilise assertions, print statements, code cell outputs, and visualisations to inform data preprocessing techniques, detect erroneous assumptions, create new features to enhance model performance, and verify hardware and software dependencies. As data, personnel, and system environments inevitably change, undocumented decisions can lead to accumulation of technical debt. This not only prolongs time required for debugging and troubleshooting, but also incurs additional costs and resource expenditures~\citep{sculley2015hidden,amershi2019software,sambasivan2021everyone}.

Our results from RQ3 reveal that implicit feedback from print and last cell statements are used in Jupyter notebooks to make design and implementation decisions throughout the machine learning development lifecycle. Unlike assertions, undocumented decisions based on implicit feedback mechanisms incur higher technical debt since they don't stop the execution despite violating the underlying assumptions of the code.

We recommend that ML practitioners document insights derived from implicit feedback mechanisms, especially for visualizations. Visual cues and trends in visualizations are open to interpretation, meaning that different practitioners may interpret the same plot differently~\citep{heer2010tour}. By documenting these visual insights, practitioners can ensure that the knowledge encapsulated in the visualizations, along with its interpretation and the subsequent decisions made, is preserved and consistently applied throughout the ML lifecycle. Although writing documentation from implicit and explicit feedback mechanisms still requires significant manual effort and resources, the dataset contributed by this study provides a valuable resource for ongoing efforts to automatically generate documentation from code written in Jupyter notebooks~\citep{yang2021subtle}.

\subsection{Automated Data Validation}

Catching data errors are critical since ML models trained using dirty data will lead to poor and incorrect predictions. This is a cause for concern since these models may be deployed in production and their output may be used by other services which will then also perform poorly. The output of the ML models are often also used to capture new training data, thus leading to a feedback loop which over time will degrade the model performance~\citep{sculley2015hidden,breck2019data}.

\citet{breck2017ml} view ML training as analogous to compilation and recommend that training data be tested, similar to code in traditional software. They propose a rubric to assess the production readiness of an ML pipeline, which includes several recommendations for data testing. In subsequent work, \citet{breck2019data} introduce \emph{TensorFlow Data Validation}, an automated tool for data validation that works on principles borrowed from the database community. Data expectations are specified in a schema, and each new data batch is validated against this schema to ensure it meets the expected criteria or prompts schema updates if requirements change.

The results from RQ2 indicate that assertions can be used to automatically validate critical assumptions in ML workflows. Embedding assertions into the code helps ensure that any deviation from expected conditions is promptly flagged. This reduces the reliance on manual verification thus enhancing the robustness and maintainability of ML pipelines. Assertions also serve as an executable form of documentation that captures the assumptions and decisions made during model development. This is particularly important in collaborative environments where multiple practitioners might work on the same project. Assertions ensure that all team members have a clear understanding of the requirements and constraints across the entire ML pipeline, facilitating smoother transitions and handovers. Moreover, assertions may enhance the reproducibility of ML experiments by ensuring that the same validation checks are applied consistently, regardless of the execution order of the notebook cells~\citep{wang2020assessing}.

Results from RQ2 and RQ3 show that assertions, print statements and last cell statements are predominantly used in the upstream, data-intensive stages of the MLDL. For instance, this study identifies several assertions that perform data validation checks to ensure data meets specific criteria or constraints. Similarly, we find assertions, print statements, and last cell statements that conduct type, shape, and distribution checks to safeguard against data inconsistencies. These results show a missed opportunity in data validation tasks. We argue that when integrating dedicated data validation tools into an operational ML pipeline, the existing feedback mechanisms in notebooks can be effectively used to write the data validation schema.

\subsection{ML Testing Literacy}

Despite the potential benefits, our results from RQ1 reveal that assertions are not widely used in Jupyter notebooks. The relatively low adoption rate compared to implicit feedback mechanisms, suggests that there is significant room for growth and improvement in the integration of assertions into ML development workflows.

We hypothesize that this limited use of assertions is due to a lack of knowledge among practitioners regarding testing approaches and the absence of adequate tooling to support the assertion of data and model properties. Guidelines and curricula on ML coding practices have historically focused on building better models. We argue that the focus must shift towards building reliable ML-enabled systems by incorporating decades of research from the software engineering community with ML. Our recommendation is evident in the recent developments within the software engineering research community that steer education and research towards addressing the specific challenges faced when developing ML-enabled software systems~\citep{kastner2020teaching}. Fostering a deeper understanding of testing practices and providing tools tailored to ML development can help promote more rigorous and reliable testing methodologies within the community.

The assertions analysed in RQ1 and RQ2 further reveal two significant challenges in the current landscape of testing in ML Jupyter notebooks. First, the assertions identified in this study lack the support of dedicated testing libraries. This suggests that the culture of testing ML code is still in its infancy, and ML practitioners are not yet fully exploring or leveraging software testing libraries. Secondly, existing testing libraries are not specifically tuned to the needs of ML code, leading practitioners to rely on basic built-in approaches. This highlights the necessity for developing specialized testing frameworks that cater to the unique requirements of ML projects.

\subsection{Machine Learning Bugs}

The findings of this study align with prior research on real-world defects in ML systems. \citet{humbatova2020taxonomy} propose a taxonomy of faults in DL systems, while \citet{morovati2023bugs} curate \emph{Defects4ML}, a dataset consisting of ML bugs. Both studies identify defects arising from the use of incorrect data types and shapes, deprecated APIs, and missing GPU resources. Our study identifies several explicit assertions that can catch such data errors. Additionally, we find several implicit checks performed using print and last cell statements to verify the presence of compute resources and libraries, monitor model training, and check the architecture of neural networks.

This study further reveals several avenues for research. We primarily focused on identifying feedback mechanisms used when writing ML code in Jupyter notebooks and subsequently exploring a limited subset of them. Further research is required to explore the feedback mechanisms in more detail and organize the knowledge into a taxonomy based on specific ML validation tasks. Future work can focus on mapping the feedback mechanisms in Jupyter notebooks to the ML defects identified by \citet{humbatova2020taxonomy} and \citet{morovati2023bugs} which can aid practitioners in debugging or preventing faults using the appropriate feedback technique.

\subsection{Code Quality and Conformance to Standards}

Results from RQ1 show that 68\% of the assertions do not contain a failure message. This can lead to problems such as unclear error diagnostics, difficulty in debugging, and increased time to identify and resolve issues. \citet{vidoni2021evaluating} investigated the testing culture in open-source R packages and identified several anti-patterns that lead to the accumulation of testing technical debt. The dataset contributed by this study can be used by researchers to replicate the \citet{vidoni2021evaluating} study using assertions written in Python.

\section{Threats to Validity}\label{sec:threats}

\paragraph{\textbf{Internal Validity}} This study examines Jupyter notebooks written in Python, specifically focusing on machine learning code. We do this by only considering notebooks that utilize popular ML libraries. Consequently, we may miss notebooks that are ML-related but do not use these specific libraries. Additionally, notebooks that import popular ML libraries but do not actively use them in the code could lead to misclassification. To mitigate this threat, we specifically look for ML libraries that facilitate model training rather than more general-purpose libraries such as pandas, numpy, and scipy, which can be used for a variety of non-ML tasks.

We enforced a time resource limit of 200 hours for conducting the case studies in RQ2 and RQ3, potentially constraining the depth and breadth of our investigation. The stratified random sampling used for selecting the case study candidates mitigates this threat by ensuring that a representative cross-section of the population is sampled within the time constraint. While this sampling method enhances the generalizability of our results, the time limitation should still be considered when interpreting and applying the study's findings.

The case studies conducted in RQ2 and RQ3 are subject to bias from the first author's individual perspective and understanding which could limit the interpretation of data and potentially overlook alternative viewpoints or explanations. To mitigate this threat, we implemented a rigorous review process involving multiple rounds of discussion with the second and third authors. These discussions continued until a consensus was reached on the interpretations and findings.

\paragraph{\textbf{External Validity}} Our analysis is based on ML notebooks publicly available on GitHub and Kaggle. Consequently, the findings may not be generalizable to Python notebooks that do not concentrate on ML. Additionally, the results may not be applicable to notebooks authored in other statistical programming languages (such as R or Julia), or to proprietary notebooks hosted on different platforms.

\section{Related Work}\label{sec:related}

\subsection{Assertions and Software Bugs}

\citet{kochhar2017revisiting} conducted a partial replication of a prior study on assertion usage, expanding the analysis to 185 Java projects from GitHub. \citet{vidoni2021evaluating} conducted a comprehensive study on the unit testing practices within 177 systematically selected open-source R packages hosted on GitHub. The research addressed the quality of testing, identified testing goals, and pinpointed potential sources of testing technical debt.

\citet{humbatova2020taxonomy} developed a comprehensive taxonomy of faults in deep learning systems through the manual analysis of 1059 artefacts from GitHub and Stack Overflow, complemented by structured interviews with 20 researchers and practitioners. The study categorized faults into five top-level categories, encompassing model, data, training, deployment, and infrastructure-related faults. \citet{morovati2023bugs} addressed the challenge of assessing the quality and performance of tools designed to improve the reliability of machine learning components by creating a standard fault benchmark called defect4ML. This benchmark consists of 100 reproducible bugs reported by ML developers on GitHub and Stack Overflow, focusing on the TensorFlow and Keras ML frameworks.

In contrast to prior work, our research focuses on the feedback mechanisms used within Jupyter notebooks for ML development. While previous work has provided insights into assertion usage and fault taxonomies, our study specifically examines the role of implicit and explicit feedback in ensuring data and model integrity within interactive development environment provided by Jupyter notebooks.

\subsection{Computational Notebooks and Software Engineering}\label{sec:notebooks}

\citet{psallidas2019data} provide an overview of the evolving landscape of computational notebooks by analysing six million Python notebooks, two million enterprise Data Science (DS) pipelines, source code and metadata from over 900 releases of 12 important DS libraries. Their findings can be used by system builders for improving DS tools and also by DS practitioners to understand the current trends in technologies they should focus on. \citet{pimentel2019large-scale} mined 1.4 million notebooks from GitHub to conduct an empirical study on the coding practices in computational notebooks. Based on their analysis, the authors propose guidelines on improving reproducibility of computational notebooks. To enable future research on computational notebooks, \citet{quaranta2021kgtorrent} mine Kaggle and present \textit{KGTorrent}, a public dataset consisting of approximately 250 thousand Jupyter notebooks written in Python. The dataset also contains a relational database dump of metadata regarding publicly available notebooks on Kaggle.

Studies with human subjects have been conducted to gain a deeper understanding of the challenges faced by ML practitioners when developing ML models inside notebooks. The results indicate that ML practitioners work in a highly iterative fashion, often experimenting with multiple strategies to analyse the data or produce meaningful visualizations~\citep{kandel2012enterprise,kery2018story,liu2019understanding,chattopadhyay2020whats}. Studies have also been conducted to understand how practitioners generate, evaluate and manage alternative hypothesis, visual designs, methods, tools, algorithms and data sources to arrive at the final implementation~\citep{liu2019understanding,kandel2012enterprise}. The findings from these studies can be used as guidelines for improving existing notebook technologies or designing new ones.

To manage and prune multiple versions of code that accumulate over time when developing in notebooks, \citet{head2019managing} propose a code gathering tool that allows practitioners to review and only keep the relevant version of code. Other tools such as \textit{WrangleDoc} and \textit{VizSmith} have also been proposed to aid ML practitioners when working within computational notebooks~\citep{yang2021subtle,bavishi2021vizsmith}.

Jupyter notebooks have been widely adopted by the DS and ML communities to develop ML pipelines~\citep{wang2020assessing,pimentel2019large-scale,quaranta2021kgtorrent}. Computational notebooks provide a rich source of data in the form of natural text, code and visualizations. Computational notebooks also allow ML practitioners to present the knowledge gained from the analysis in a narrative that can be shared with others~\citep{rule2018exploration}.

Our study builds on this foundation by focusing on the specific feedback mechanisms used within Jupyter notebooks during ML development. We examine both implicit and explicit feedback methods, providing a detailed analysis of their usage and effectiveness in maintaining robust ML workflows.

\subsection{ML Testing}\label{sec:ml-testing}

Existing scientific literature on ML testing broadly focuses on two aspects. First, on functional properties such as correctness and robustness of the model towards unseen data. And second, on non-functional properties such as fairness, interpretability and privacy~\citep{zhang2022machine,mehrabi2021survey,chen2022fairness}.

To test the correctness of ML models, several improvements over existing test adequacy metrics have been proposed. Tools such as \textit{DeepXplore} and \textit{DeepGauge} propose new test adequacy metrics such as neuron coverage adapted for ML enabled software systems~\citep{pei2017deepxplore,ma2018deepgauge,gerasimou2020importance-driven}. Formal verification methods have also been proposed that try to provide formal guarantee of robustness against adversarial examples. Such methods however are only feasible for statistical ML models and become computationally expensive for more complex models such as deep neural networks~\citep{zhu2021deepmemory,baluta2021scalable}. Several studies have been conducted on generating and detecting adversarial inputs for ML models. Data augmentation techniques based on fuzzing, search based software testing and mutation testing have been proposed to generate adversarial examples that can be used during model training to improve its robustness~\citep{braiek2019deepevolution,gao2020fuzz,wang2021robot,zhang2020white-box}. It is however not possible to include all variations of adversarial examples into the training data. Thus, methods have been proposed to detect adversarial inputs during runtime~\citep{xiao2021self-checking,wang2020dissector,wang2019adversarial,berend2020cats}.

Despite these advancements, there is a notable gap in practical, day-to-day testing practices within ML development environments. Our study addresses this gap by analysing the feedback mechanisms in Jupyter notebooks, highlighting how assertions and print statements are used to validate data and model performance, and proposing ways to integrate these practices more effectively into the ML development lifecycle.

\subsection{Visualizations and Machine Learning}\label{sec:visualizations}

As ML augment software systems in safety-critical domains, emphasis has been put into explainability of ML models. ML models and the underlying data is complex and multidimensional. To combat the ``curse of dimensionality'', visual analytics has been widely adopted by the ML community to understand the data and the internal workings of ML models~\citep{yuan2021survey,hohman2019visual,wexler2019what-if}.

Prior studies have been conducted to understand how visual analytics techniques are currently being used in ML. \citet{yuan2021survey} conduct a systematic review of 259 papers and propose a taxonomy of visual analytics techniques for ML. \citet{hohman2019visual} conduct a survey of visual analytics techniques for Deep Learning Models. The findings of the study indicate that visual analytics has been widely adopted in ML for model explanation, interpretation, debugging and improvement.

Several tools have been proposed by the visual analytics research community to aid practitioners in understanding how their ML models operate. \citet{wexler2019what-if} propose \textit{What-If}, a visual analytics tool to explore alternative hypotheses, generate counterfactual reasoning, investigate decision boundaries of the model and how change in data affects the model predictions. To reduce the cognitive load of ML practitioners during model building phase, \citet{amershi2019software} propose \textit{ModelTracker}. Given a trained model and a sample dataset, ModelTracker presents all the information in traditional summary statistics along with the performance of the model.

\textit{ESCAPE}, \textit{GAM Coach}, \textit{Angler} and \textit{Drava} are a few other tools that have been proposed to handle specialized use-cases such as identifying systematic errors in ML models, generating counterfactual explanations for Generalized Additive Models, prioritizing machine translation model improvements and relating human concepts with semantic dimensions extracted by ML models during disentangled representation learning~\citep{ahn2023escape,wang2023gam,robertson2023angler,wang2023drava}.

Unlike prior work that focuses on standalone visual analytics tools, our study explores how visualizations within Jupyter notebooks are used by ML practitioners. We analyse the motivations behind creating these visualizations and the insights they provide, offering a nuanced understanding of their role in the iterative ML development process.

\section{Conclusion}

This study provides a comprehensive analysis of feedback mechanisms employed within Jupyter notebooks during the machine learning development lifecycle. Our findings indicate that while implicit feedback mechanisms such as print statements and last cell statements are extensively utilized, there is a significant gap in the adoption of explicit feedback mechanisms like assertions. This reliance on manual validation highlights the potential for automated data validation to enhance the robustness and maintainability of ML pipelines. This study also underscores the necessity for better documentation practices to mitigate technical debt and ensure the reproducibility of ML experiments. Feedback mechanisms are used throughout the ML development workflow to achieve more reliable and efficent model development processes. Future work should focus on developing specialized tools and frameworks to support these practices, thereby advancing the field of ML engineering.

\bibliographystyle{spbasic}      
\bibliography{bibliography}   

\end{document}